\documentclass[twocolumn]{aastex62}

\usepackage{amsmath,amsopn,amsxtra}
\usepackage{comment} 
\usepackage{float}
\usepackage{natbib} 

\newcommand{\ha}{H$\alpha$}

\newcommand{\vlsr}{{\rm v}_{\rm LSR}}

\newcommand{\kms}{ \ifmmode{\rm km\thinspace s^{-1}}\else km\thinspace s$^{-1}$\fi}
\newcommand{\kmse}{\mbox{km s$^{-1}$}} 

\newcommand{\pc}{\ensuremath{ \, \mathrm{pc}}}
\newcommand{\kpc}{\ensuremath{\, \mathrm{kpc}}}
\newcommand{\cm}{\ensuremath{ \, \mathrm{cm}}}

\newcommand{\hi}{H\textsc{~i}}

\shorttitle{Complex~A}
\shortauthors{Barger et al.}

\begin{document}

\title{Exploring Hydrodynamic Instabilities along the Infalling High-Velocity Cloud Complex~A}

\author[0000-0001-5817-0932]{Kathleen A. Barger}
\affiliation{Department of Physics \& Astronomy, Texas Christian University, Fort Worth, TX 76129, USA}

\author[0000-0002-1793-3689]{David L. Nidever}
\affiliation{Department of Physics, Montana State University, P.O. Box 173840, Bozeman, MT 59717-3840}
\affiliation{National Optical Astronomy Observatory, 950 North Cherry Ave, Tucson, AZ 85719}

\author{Cannan Huey-You}
\affiliation{Department of Physics \& Astronomy, Texas Christian University, Fort Worth, TX 76129, USA}
\affiliation{Accommodated Learning Academy, Grapevine, TX 76051, USA}

\author[0000-0001-9158-0829]{Nicolas Lehner}
\affiliation{Department of Physics, University of Notre Dame, Notre Dame, IN 46556, USA}

\author{Katherine Rueff}
\affiliation{Department of Physics, University of Notre Dame, Notre Dame, IN 46556, USA}

\author{Paris Freeman}
\affiliation{Department of Physics \& Astronomy, Texas Christian University, Fort Worth, TX 76129, USA}
\affiliation{Founders Classical Academy of Lewisville, Lewisville, TX 75067, USA}

\author{Amber Birdwell}
\affiliation{Department of Physics \& Astronomy, Texas Christian University, Fort Worth, TX 76129, USA}
\affiliation{Aledo High School, Aledo, TX 76008, USA}

\author{Bart P. Wakker}
\affiliation{Supported by NASA/NSF, affiliated with Department of Astronomy, University of Wisconsin-Madison, Madison, WI 53706, USA}

\author{Joss Bland-Hawthorn}
\affiliation{Sydney Institute for Astronomy, School of Physics A28, University of Sydney, NSW 2006}

\author{Robert Benjamin}
\affiliation{University of Whitewater, Whitewater, WI 53190, USA}

\author[0000-0002-1295-988X]{Drew A. Ciampa}
\affiliation{Department of Physics \& Astronomy, Texas Christian University, Fort Worth, TX 76129, USA}

\begin{abstract}

Complex~A is a high-velocity cloud that is traversing through the Galactic halo toward the Milky Way's disk. We combine both new and archival Green Bank Telescope observations to construct a spectroscopically resolved H\textsc{~i}~21-cm map of this entire complex at a $17.1\lesssim\log({N_{\rm H\textsc{~i},\,1\sigma}}/\cm^{-2})\lesssim17.9$ sensitivity for a ${\rm FWHM}=20~\kms$ line and  $\Delta\theta=9\farcm1$ or $17\lesssim\Delta d_{\theta}\lesssim30~\rm pc$ spatial resolution. We find that that Complex~A is has a Galactic standard of rest frame velocity gradient of $\Delta\rm v_{GSR}/\Delta L=25~\kms/{\kpc}$ along its length, that it is decelerating at a rate of $\langle a\rangle_{\rm GSR}=55~{\rm km}/{\rm yr}^2$, and that it will reach the Galactic plane in $\Delta t\lesssim70~{\rm Myrs}$ if it can survive the journey. We have identify numerous signatures of gas disruption. The elongated and multi-core structure of Complex~A indicates that either thermodynamic instabilities or shock-cascade processes have fragmented this stream. We find Rayleigh-Taylor fingers on the low-latitude edge of this HVC; many have been pushed backward by ram-pressure stripping. On the high-latitude side of the complex, Kelvin-Helmholtz instabilities have generated two large wings that extend tangentially off Complex~A. The tips of these wings curve slightly forward in the direction of motion and have an elevated \hi\ column density, indicating that these wings are forming Rayleigh-Taylor globules at their tips and that this gas is becoming entangled with unseen vortices in the surrounding coronal gas. These observations provide new insights on the survivability of low-metallicity gas streams that are accreting onto $L_\star$ galaxies.

\end{abstract}


\section{Introduction}\label{section:intro}

The star formation in galaxies is dependent on their ability to accrete gas onto their disks. Although both the Milky Way and Andromeda are surrounded by gas (e.g., \citealt{1997ARA&A..35..217W,2003ApJS..146....1W,2004A&A...417..421B, 2011Sci...334..955L, 2015ApJ...804...79L}), their star-formation rates appear to be in a decline (see  \citealt{2016ARA&A..54..529B} for a review). These galaxies may even be transitioning into the ``Green Valley'' \citep{2011ApJ...736...84M, 2012ApJ...751...74D, 2016ARA&A..54..529B}, which is the region between blue star-forming and red quiescent galaxies on a color-magnitude diagram.
    
The halo clouds that surround the Milky Way are typically put into two different categories that are based on their local standard of rest (LSR) velocities. Intermediate-velocity clouds (IVCs) are a slower population ($30\lesssim|\rm v_{LSR}|\lesssim90~\kms$) that tend to lie near the Galactic disk. 
The high-velocity cloud (HVC) population ($|\rm v_{LSR}|\gtrsim90~\kms$) has multiple origins, including galactic-feedback processes, halo-gas condensations, nearby low-mass galaxies, and intergalactic medium filaments; therefore, many HVCs provide replenishment the star-formation reservoir of our galaxy (\citealt{2012ARA&A..50..491P} and \citealt{2017ASSL..430...15R} for review).   

As HVCs clouds travel through galaxy halos, they heated and ionized by photons that are escaping from the galaxies (e.g., Milky Way: \citealt{1999ApJ...510L..33B,2001ApJ...550L.231B, 2005ApJ...630..332F}; Magellanic Clouds: \citealt{2013ApJ...771..132B}). Additionally, the hot coronal gas that surrounds them acts as a headwind that compresses their leading material and strips its outer layers through a process known as ram-pressure stripping (e.g., \citealt{2011MNRAS.418.1575P, 2014ApJ...792...43F}). When the surrounding gas rubs against the HVC's surface, it promotes Kelvin-Helmholtz instabilities---a type of shear-driven disturbance---which can cause small cloudlets to fracture off the complex's main body (see \citealt{2007ApJ...671.1726S,2007ApJ...670L.109B, 2009ApJ...698.1485H}). Rayleigh-Taylor instabilities, which are buoyancy-driven disturbances, further disrupt the complex because it is resting on top of less dense halo gas while situated in a galaxy's gravitational field. Combined, these processes can cause the skin of the cloud to become warmer, ionized, and more diffuse than its core. Internal temperature and density variations between these two gas phases can generate thermal instabilities, which can fragment the cloud (see \citealt{2004ApJ...615..586M}). Fragmentation can also occur if stripped leading gas, due to ram-pressure stripping, collides with down stream material (see \citealt{2007ApJ...670L.109B, 2015ApJ...813...94T}). As the surface area of the HVC increases, it becomes more exposed to its environment, which will cause it to  evaporate more rapidly into the surrounding coronal gas (e.g., \citealt{2002A&A...391..713K}). 

Complex~A is plummeting towards the Galactic disk and could supply our galaxy with up to $M_{\rm total}\gtrsim2\times10^6~\rm M_\odot$ (neutral: \citealt{1994A&A...282..709K,2004ASSL..312..195V}; ionized: \citealt{2012ApJ...761..145B}) of new material (${Z}=0.1~{ Z}_\odot$: \citealt{1994A&A...282..709K, 1995A&A...302..364S, 1999Natur.400..138V, 2001ApJS..136..463W,2012ApJ...761..145B}). Its chemical composition indicates that it either originated from a low-mass galaxy or the intergalactic medium. However, as no complementary stellar stream has been found \citep{2010ApJ...712L.103B,2010ApJ...711...32N}, Complex~A was not likely stripped from a satellite galaxy. This complex has an elongated morphology with multiple dense cores---dubbed A0--AVI and B---along its $\Delta L\approx6.4~\kpc$ length ($\Delta\theta\approx35\arcdeg$; \citealt{2012ApJ...761..145B}). Because this infalling cloud spans $3 \lesssim z \lesssim 7~\kpc$ above the Galactic disk  (\citealt{1996ApJ...473..834W,2003ApJS..146....1W, 1997MNRAS.289..986R, 1999Natur.400..138V, 2012ApJ...761..145B, 2012MNRAS.424.2896L}), it probes a range of halo conditions that vary with height above the disk. 

In this study, we investigate how HVC Complex~A is affected by its environment with new and archival Green Bank Telescope (GBT) H\textsc{~i}~21-cm observations. We describe these observations and their reduction in Sections~\ref{section:obs} and~\ref{section:reduction}. We outline our Gaussian decomposition procedure in Section~\ref{section:gauss_decoms} and explore the H\textsc{~i} morphology and kinematic structure along the length of the complex in Section~\ref{section:kinematics_morphology} and discuss morphological features that are indicative of hydrodynamic instabilities occurring within different regions of Complex~A. Finally, we summarize our main conclusions in Section~\ref{section:summary}. 

\section{Observations}\label{section:obs}

\begin{figure}
\begin{center} 
\includegraphics[trim=0 0 0 0,clip,scale=0.23,angle=0]{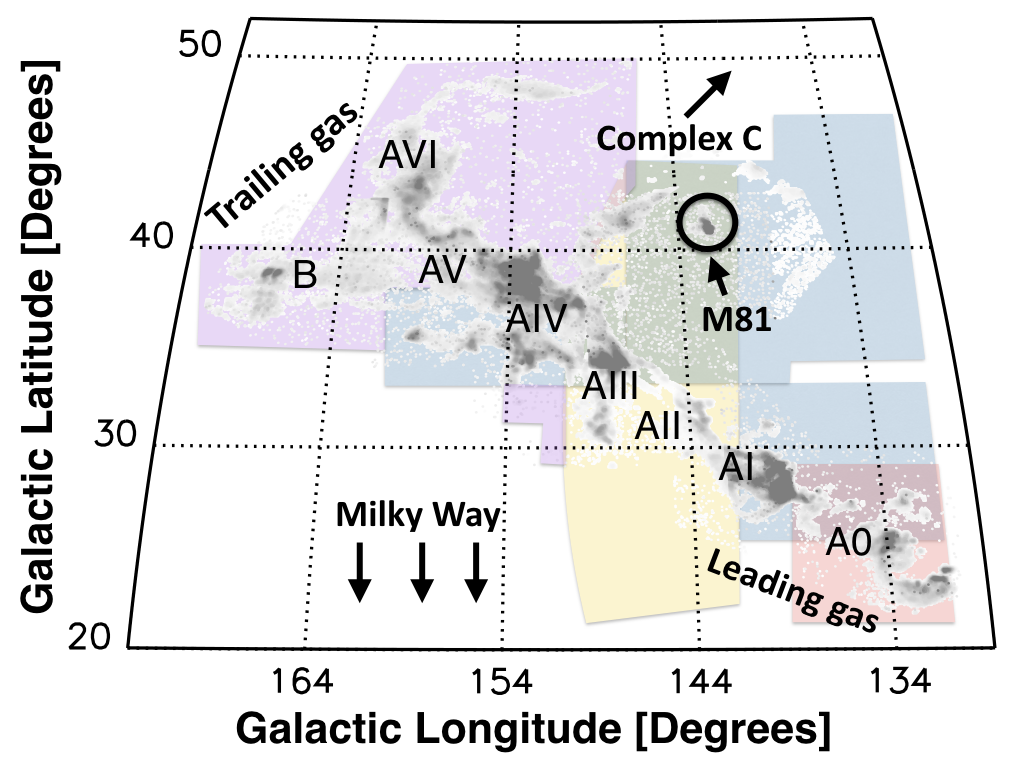} \\
\end{center}
\figcaption{An H\textsc{~i} map of Complex~A with the regions that were observed through different programs highlighted (see  Table~\ref{table_data}) and with the eight high H\textsc{~i} column density cores~A0--AVI and core~B labeled. The two purple shaded regions mark the locations of our new observations that primarily span the trailing portion of this complex (PI Barger: GBT13B-068). The region highlighted in red spans the leading portion of this gaseous stream (PI Verschuur: ID GBT1010A-003). The yellow (PI Chynoweth: GBT09A-046) and blue (PI Martin: GBT107A-003) regions indicate observations cover the central region of this HVC.  We additionally circle the emission in our surveyed region that is associated with the M81 galaxy at $(l,~b)=(142\fdg1,~40\fdg9)$.
\label{figure:observed_regions}}
\end{figure}

\begin{deluxetable*}{lccccccc}
\tablecaption{GBT Data}
\tablecolumns{8}
\tablewidth{0pt}
\tablehead{
\colhead{Prop ID} & \colhead{PI} & \colhead{Name} & \colhead{$l,~b$ Center} & \colhead{Angular Size} & \colhead{$\Delta\theta$} & \colhead{$\Delta\rm v_{LSR}$} & \colhead{$\log({N_{\rm H\textsc{~i},\,1\sigma}}/\cm^{-2})$\tablenotemark{a}} \\
\colhead{} & \colhead{} & \colhead{} & \colhead{(degrees)} & \colhead{(degrees)} & \colhead{(arcmins)} & \colhead{(km s$^{-1}$)} & \colhead{}
}
\startdata
GBT09A-046\tablenotemark{b} & Chynoweth & M81 & $143\fdg0,~35\fdg2$  & 9\degr$\times$24\degr & $1\farcm75$ &  5.2 & $17.1$ \\
GBT10A-003 & Verschuur & A0 & $134\fdg2~,25\fdg1$ & 8.5\degr$\times$7\degr & $3\farcm5$ &  0.48 & $17.6$ \\
GBT07A-104 & Martin & UMA & $144\fdg3,~38\fdg6$ & 9\degr$\times$9\degr & $3\farcm5$ &  0.80 & $17.7$ \\
GBT07A-104 & Martin & UMAEAST & $155\fdg8,~37\fdg0$ & 10.5\degr$\times$6\degr & $3\farcm5$ &  0.80 & $17.8$ \\
GBT07A-104\tablenotemark{c} & Martin & SPC  & $135\fdg4,~30\fdg3$ & 12\degr$\times$9.5\degr & $3\farcm5$ &  0.80 & $17.9$ \\
GBT13B-068 & Barger & B & $166\fdg0,~39\fdg0$ & 9\degr$\times$7.5\degr &  $3\farcm5$ &  0.16 & $17.7$ \\
GBT13B-068 & Barger & AV & $155\fdg1,~41\fdg3$ & 8.5\degr$\times$3\degr &  $3\farcm5$ &  0.16 & $17.7$ \\
GBT13B-068 & Barger & AVI & $155\fdg3,~46\fdg2$ & 12.5\degr$\times$7.5\degr &  $3\farcm5$ &  0.16 & $17.7$ \\
GBT13B-068 & Barger & AIV & $153\fdg2,~33\fdg0$ & 4.5\degr$\times$3\degr &  $3\farcm5$ &  0.16 & $17.7$ \\
GBT13B-068 & Barger & AIVSOUTH & $152\fdg2,~29\fdg7$ & 2\degr$\times$1.5\degr &  $3\farcm5$ &  0.16 & $17.7$
\enddata
\tablenotetext{a}{The 1-sigma sensitivity for a ${\rm FWHM}=20~\kms$ line. A sensitivity map of all the mapped observations explored in this study is presented in Figure~\ref{figure:noise_map}.}
\tablenotetext{b}{This program scanned in ICRS coordinates while all others used Galactic coordinates.}
\tablenotetext{c}{This region also includes data from GBT08B-083 (PI: Goncalves), GBT08B-038 (PI: Goncalves) and GBT10A-078 (PI: Lockman).}
\label{table_data}
\end{deluxetable*}

\begin{figure}[t]
\begin{center}
 \includegraphics[width=1.0\hsize,angle=0]{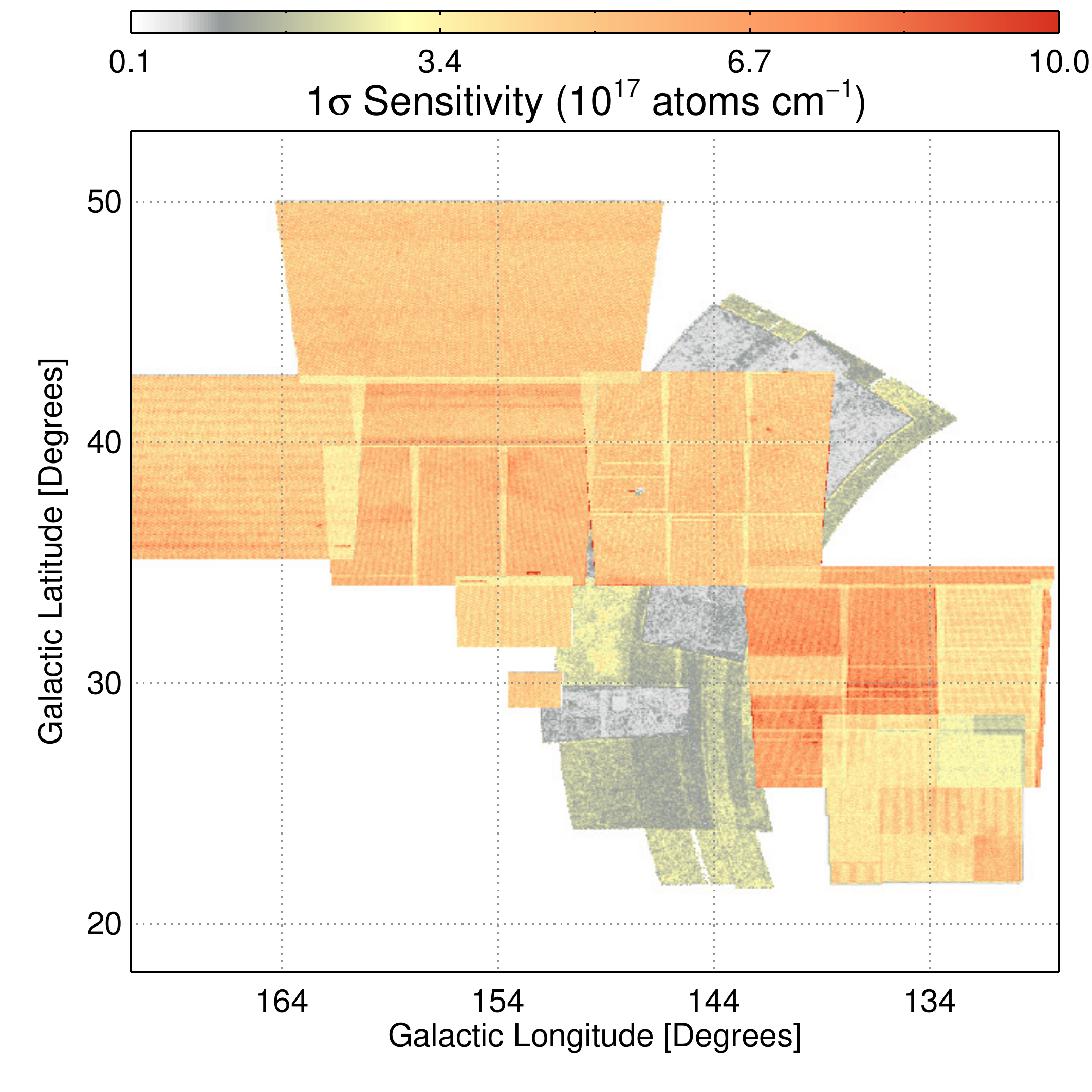}
\end{center}
\figcaption{Sensitivity map (1$\sigma$) of the Complex~A GBT observations in units of 10$^{17}$ atoms cm$^{-2}$ over 20 \kmse.}
\label{figure:noise_map}
\end{figure}

\begin{figure*}
\begin{center} 
\includegraphics[trim=0 30 0 0 0,clip,scale=0.5,angle=0]{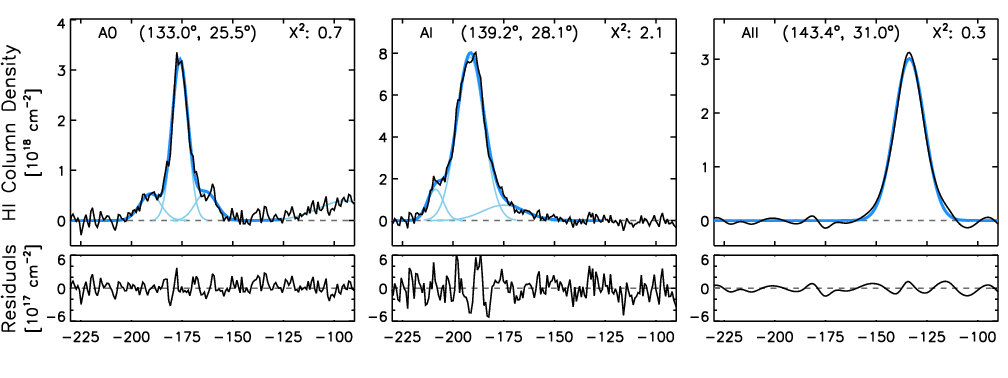} \\
\includegraphics[trim=0 30 0 0,clip,scale=0.5,angle=0]{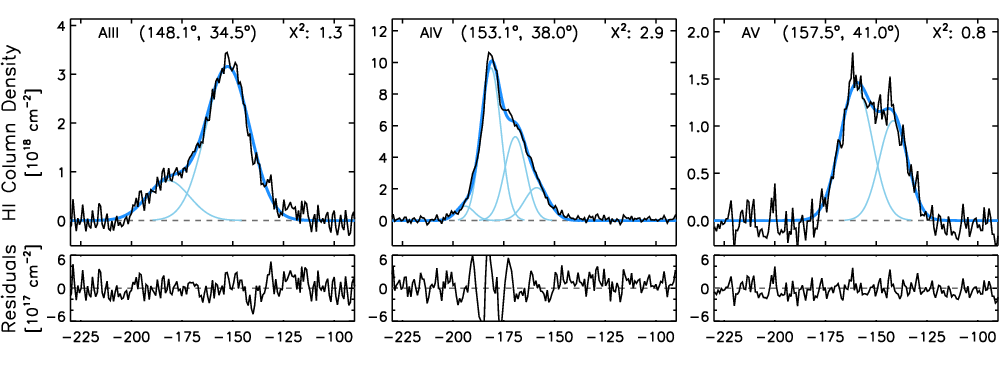} \\
\includegraphics[trim=0 0 0 0,clip,scale=0.5,angle=0]{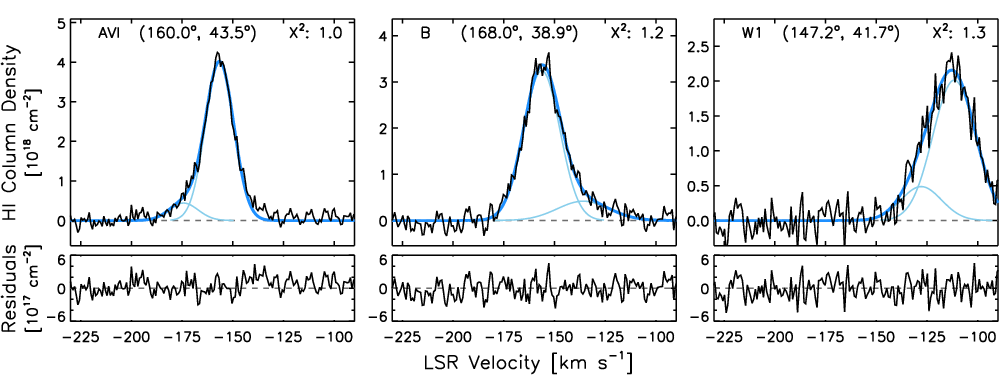} 
\end{center}
\figcaption{Example H\textsc{~i} 21-cm spectra along cores~A0--B and a high-latitude wing that is adjacent to core~AVI, which we labeled ``W1''. We include the Galactic coordinates of each spectra and the reduced Chi-squared for our fits at the top of the spectral figures. The light blue Gaussian emission-line profiles represent the component solutions to our Gaussian decompositions routine that is described in Section~\ref{section:gauss_decoms} and the dark blue traces the total fit. The bottom panels display the residuals of our fits (see Equation~\ref{eq:residuals}).
\label{figure:spectra}}
\end{figure*}

\begin{figure}
\begin{center} 
\includegraphics[trim=0 0 0 0,clip,scale=0.45,angle=0]{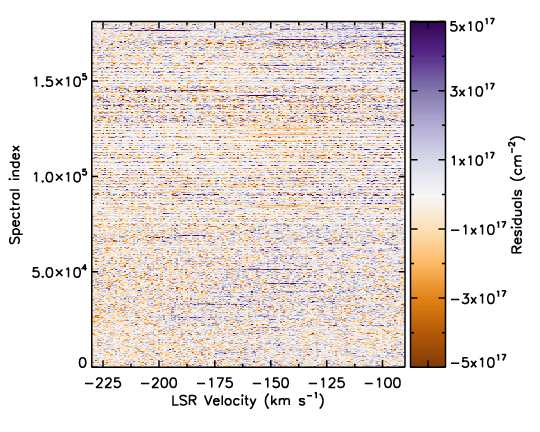} \\
\end{center}
\figcaption{Vertically stacked residuals image for each of the $\sim1.2\times10^5$ sightlines explored in this study as a function of velocity, where the residuals are defined as the difference between the H\textsc{~i} observations and the Gaussian decomposition fit. We exclude a small region of our survey shown in Figure~\ref{figure:all_regions} that is contaminated by emission from the M81 galaxy system contained between $141\arcdeg\lesssim l\lesssim 144\arcdeg$ and $40.5\arcdeg\lesssim b\lesssim 43\arcdeg$. In the bottom right hand corner, the column density of the residuals increases where core~A0 overlaps with the Milky Way; to avoid confusion with our Galactic disk, we only report on the properties of the H\textsc{~i} emission below $\rm v_{LSR}\le-130~\kms$ in this region.
\label{figure:residuals}}
\end{figure}

\begin{figure*}[t]
\begin{center}
 \includegraphics[trim=0 52 0 0,clip,scale=0.3,angle=0]{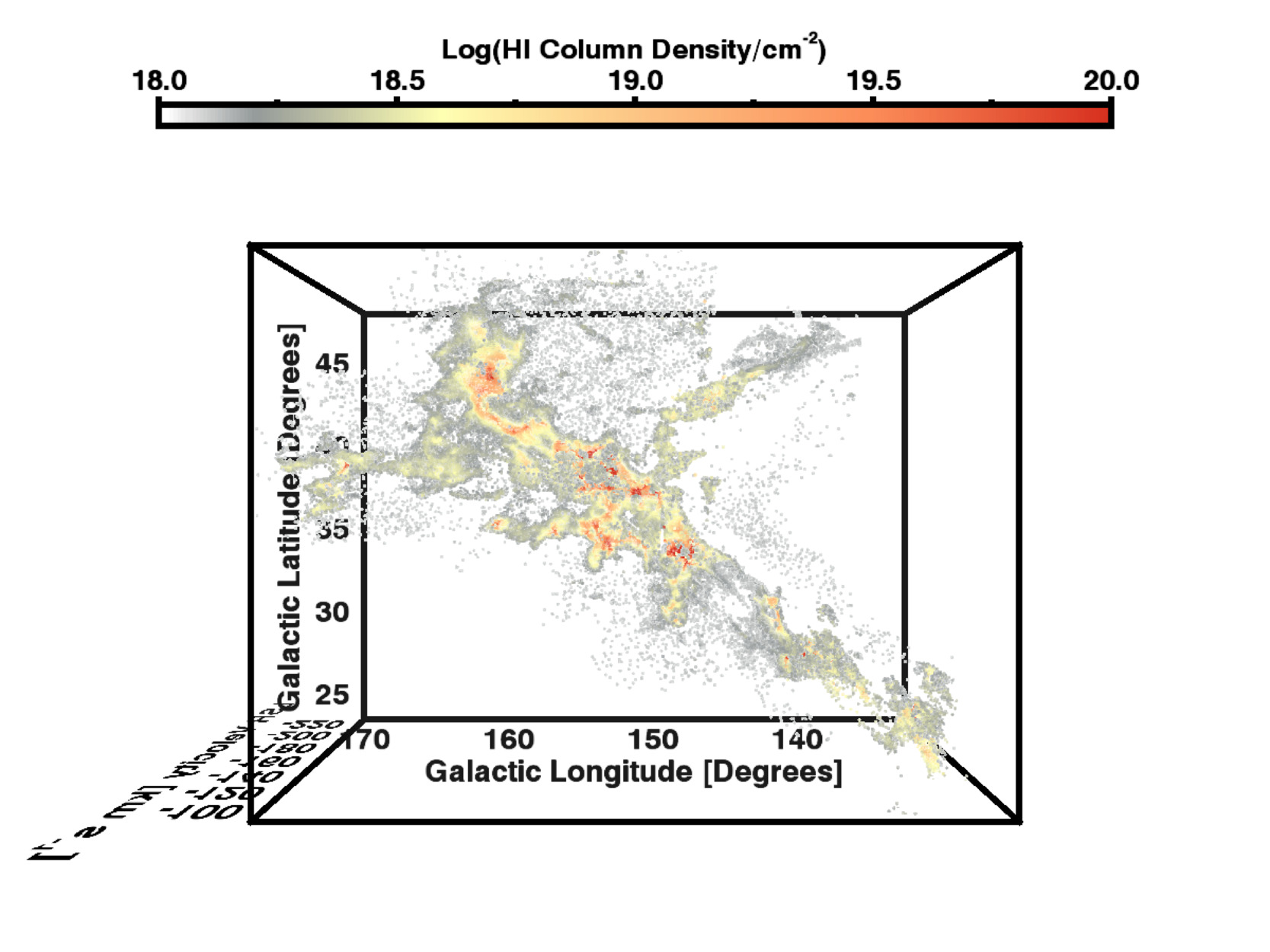} 
\end{center}
\figcaption{A 3~dimensional movie of Complex~A's \hi\ gas distribution that rotates through position-position-velocity maps. The H\textsc{~i} column densities and LSR line center are solutions to our Gaussian decomposition fits (see Section~\ref{section:gauss_decoms}). All points in the 3D map have a transparency of 50\%. The movie begins with an H\textsc{~i} position-position map that is rotated by 90$\arcdeg$ about the Galactic Longitude axis into a Galactic longitude position-velocity map. The 3D map in this movie is then rotated by another 360$\arcdeg$ about the Galactic Latitude and Longitude plane, passing through Galactic longitude position-velocity and Galactic latitude position-velocity maps twice along the way. Finally, the 3D map is rotated back into its original position by rotating 90$\arcdeg$ about the Galactic Longitude axis. The real-time duration of the movie is 1~minute and 21~seconds.
\label{figure:3d_decompositions}}
\end{figure*}

Our H\textsc{~i}~21-cm emission-line survey of Complex~A spans a $600$-square degree area across the sky. This survey is composed of new and archival 100-m Robert C. Byrd Green Bank Telescope (GBT) observations that are spectroscopically resolved over the $-230\le\rm v_{LSR}\le -90~\kms$ velocity range\footnote{Throughout this study, we use the kinematic definition of the LSR, where the solar motion moves at $20~\kms$ toward $(\rm R.A.,~DEC.)_{\rm J2000}=(18^h 3^m 50.29^s, 30\arcdeg00\arcmin16\farcs8)$.} and are spatially resolved at  $\Delta\theta=9\farcm1$ or $17\lesssim\Delta d_{\theta}\lesssim30~\rm pc$ at the distance of Complex~A ($6.3\lesssim d_{\odot}\lesssim11.3~\kpc$: \citealt{2012ApJ...761..145B}). The upper velocity limit of this survey ($\rm v_{LSR}\le -90~\kms$) is to avoid contributions from the Milky Way's disk and to reduce the contribution from the neighboring HVC Complex~C (see Figure~\ref{figure:observed_regions}). For the core~A0 region, which lies nearest to the Galactic disk, we truncated this velocity limit to $\rm v_{LSR}\ge -130~\kms$ in an effort to avoid Milky Way contamination. It is important to note that we did not search for H\textsc{~i} emission associated with Complex~A below a Galactic latitude of $b<21\fdg5$ as confusion with the Milky Way becomes to great.

\begin{figure*}
\begin{center} 
\includegraphics[trim=0 90 0 50,clip,scale=0.46,angle=0]{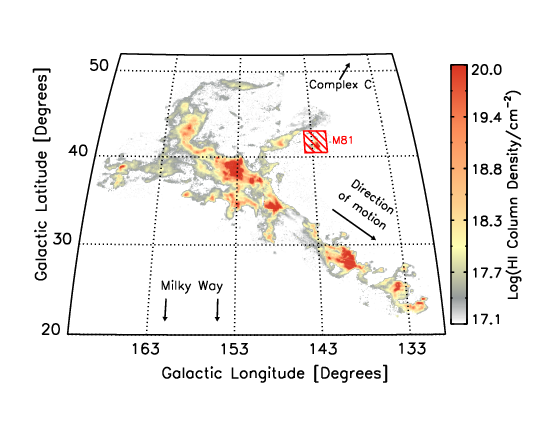}\includegraphics[trim=0 90 0 50,clip,scale=0.46,angle=0]{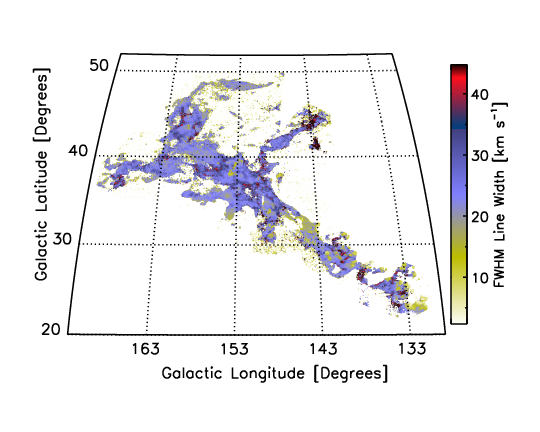}  \\
\includegraphics[trim=0 40 0 50,clip,scale=0.46,angle=0]{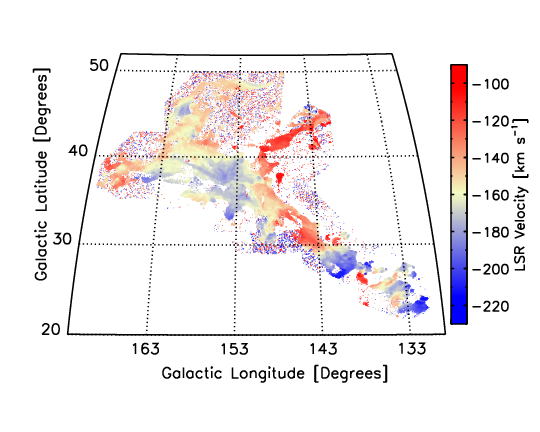}\includegraphics[trim=0 40 0 50,clip,scale=0.46,angle=0]{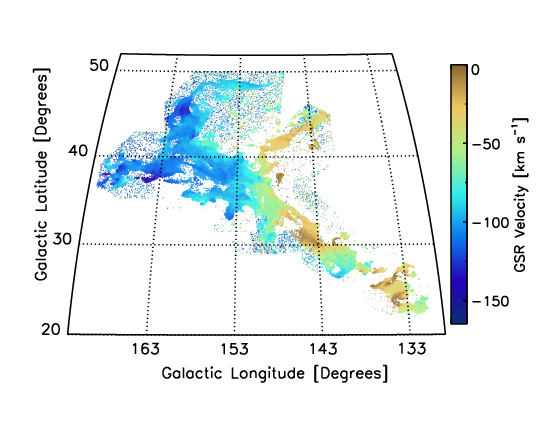}  \\
\end{center}
\figcaption{Gas distribution and motions of Complex~A over a $-230\le\rm v_{LSR}\le-90~\kms$ velocity range for the entire complex except in the core~A0 region, which is over a $-230\le\rm v_{LSR}\le-130~\kms$ velocity range to avoid emission associated with the Galactic disk. Top Left: \hi\ column density; Top Right: FWHM of the emission line; Bottom Left: Center LSR velocity of the emission line; Bottom Right: Center GSR velocity of the emission line. All values in these three maps depict solutions from Gaussian decompositions along each sightline, where the shown color represents the properties of the component with the highest \hi\ column density.  
\label{figure:all_regions}}
\end{figure*}

Our new observations from program GBT13B-068 (PI Barger) span more than a $175$~square degree region across the sky and survey the trailing half of Complex~A (see Figure~\ref{figure:observed_regions} and Table~\ref{table_data}). Each individual observation had a $4$~second exposure time for $50.8~{\rm hours}$ of integrated on-target time for all 45,695 sightlines. We centered these L-Band ($1.15\le\nu_{\textrm{L-Band}}\le1.73~{\rm GHz}$) observations on the H\textsc{~i}~21-cm line ($\nu=1420.4~{\rm MHz}$) and took them in the on-the-fly (OTF) spectral-line mapping mode. These observations span a bandwidth of $\Delta \nu =12.5~{\rm MHz}$, which corresponds to 16,384~channels that have a $\Delta\rm v_{\rm channel}=0.0583~\kms$ channel width. 

The archival GBT H\textsc{~i} 21-cm observations sample (1) cores~AII and~AIII over a $\sim$215 deg$^2$ region on the sky (PI Chynoweth: GBT09A-046), (2) core~A0 over a $\sim$60 deg$^2$ region (PI Verschuur: GBT10A-003), and (3)  cores~AI, AIII, AIV, and~AV over a $\sim$260 deg$^2$ region (PI Martin: GBT07A-104, GBT08A-083, and GBT10A-078).  Table~\ref{table_data} summarizes the angular extent, angular and spectral resolution, and the sensitivity of these datasets. The observations from the Verschuur and Planck programs were taken with a $4$~second exposure time and the ones from the Chynoweth program were taken with $5$~second exposures. Together, these archival datasets stretch over a $470$~square degrees region across the sky along $\sim$123,000 sightlines (see Figure~\ref{figure:observed_regions}). 

\begin{figure}[t]
\begin{center}
 \includegraphics[width=1.0\hsize,angle=0]{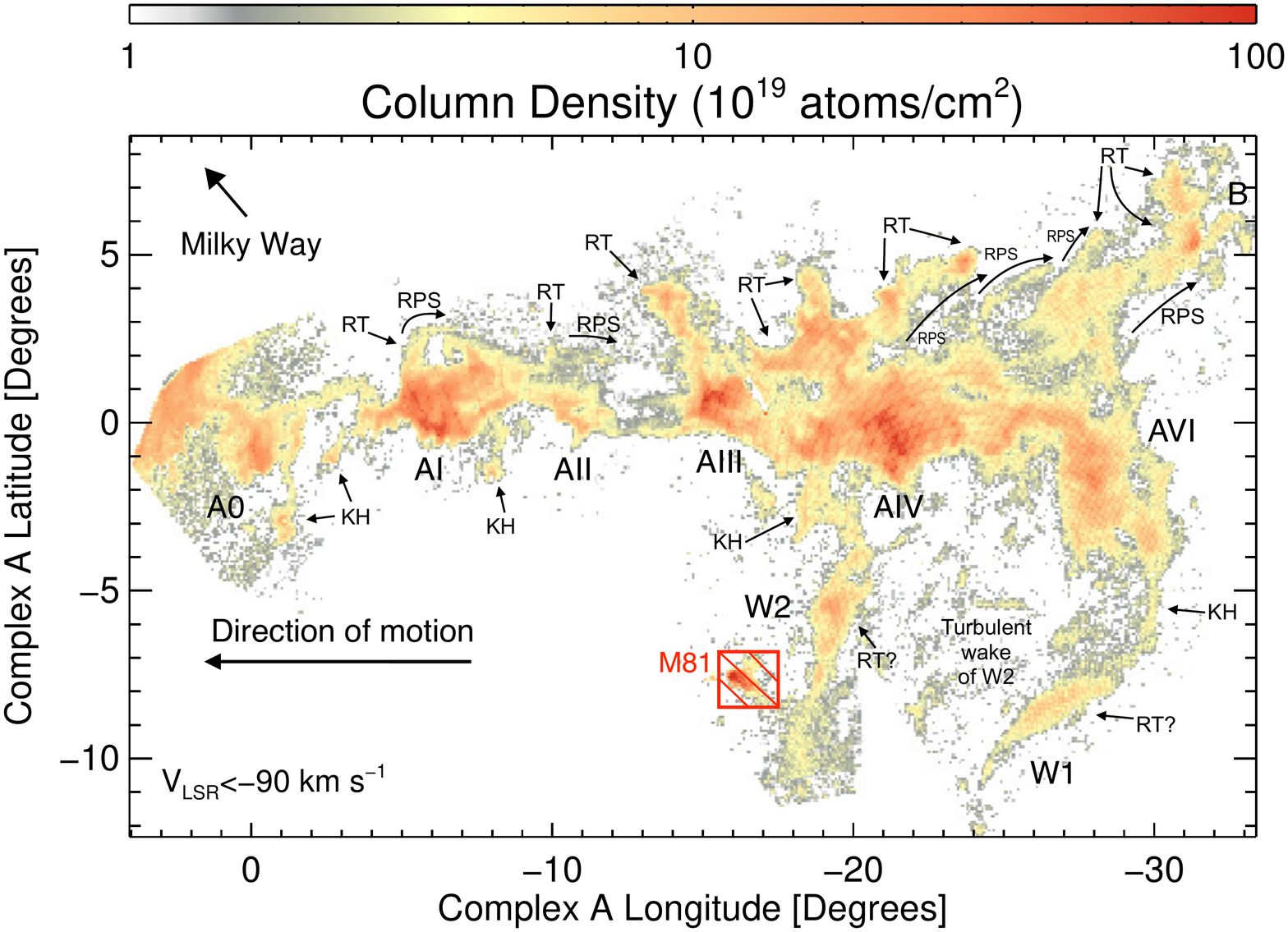}
\end{center}
\figcaption{The total integrated H\textsc{~i} column density distribution of Complex~A in the CA coordinate system, which places the main body of this HVC stream at $b_{\rm CA}=0\arcdeg$ and the center of core~A0 at $(l_{\rm CA},\,b_{\rm CA})=(0\arcdeg,\,0\arcdeg)$ or $(l,~b)=(133\fdg9,~+25\fdg1)$, where the $N_{\rm H\textsc{~i}}$ was determined using the Gaussian decompositions procedure described in Section~\ref{section:gauss_decoms}. All have marked all major cores (A0--AVI and B) are marked and high-latitude wing~1 (W1) and wing~2 (W2) that lies off the core~AVI and AIV regions, respectively. We label prominent Rayleigh-Taylor (RT) globules, Kelvin-Helmholtz structures, and ram-pressure stripping (RPS) features. The red box $(l,~b)=(142\fdg1,~40\fdg9)$ encompasses emission from the M81 galaxy.}
\label{figure:alonalat_coldens}
\end{figure}

\begin{figure}[t]
\begin{center}
 \includegraphics[width=1.0\hsize,angle=0]{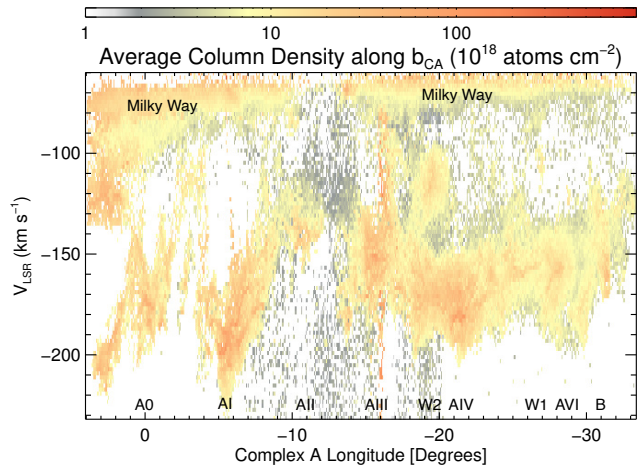}
\end{center}
\figcaption{Position-velocity map showing the column density averaged in $b_{\rm CA}$ of Complex~A and the Milky Way H\textsc{~i}~21-cm emission along $l_{\rm CA}=0\arcdeg$ in the CA coordinate System. The $N_{\rm H\textsc{~i}}$ and $\rm v_{LSR}$ center positions of the emission for each component along all sightlines was determined using the Gaussian decomposition procedure described in Section~\ref{section:gauss_decoms}.}
\label{figure:alonvlsr}
\end{figure}

\begin{figure}
\begin{center} 
\includegraphics[trim=10 25 15 10,clip,scale=0.45,angle=0]{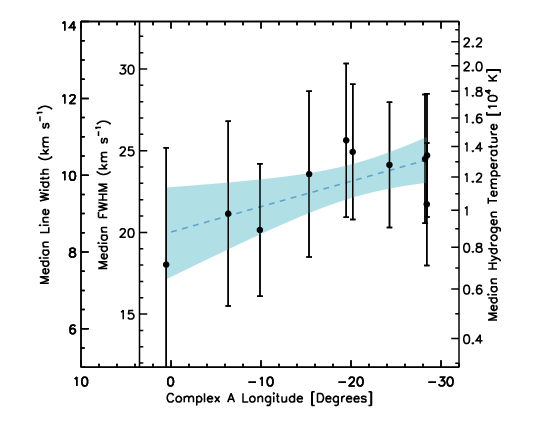}
\end{center}
\figcaption{The median line widths and hydrogen temperatures for each of the major \hi\ core regions (see Figures~\ref{figure:decomp_region_A0}--\ref{figure:decomp_region_B}) and the two high-latitude wing regions (see Figures~\ref{figure:decomp_region_W1}--\ref{figure:decomp_region_W2}). The temperatures were estimated from median width of the lines under the assumption that they are only affected by thermal-line broadening affects. The uncertainties in the line width represent the average deviation from the median. The blue dashed line marks our best fit for the ${\rm FWHM}$ as a function of Complex~A Longitude, which is given by the following linear relationship: ${\rm FWHM} = \left(-0.16\pm0.07\right)\,\kms\,{\rm degree}^{-1}\,l_{\rm CA}+\left(20.0\pm1.6\right)\,\kms$
\label{figure:clon_temp}}
\end{figure}

\begin{figure*}
\begin{center} 
\includegraphics[trim=10 25 15 10,clip,scale=0.35,angle=0]{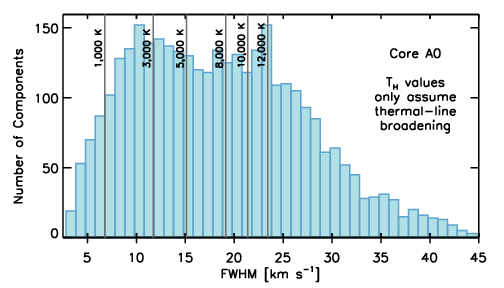}\includegraphics[trim=33 25 15 10,clip,scale=0.35,angle=0]{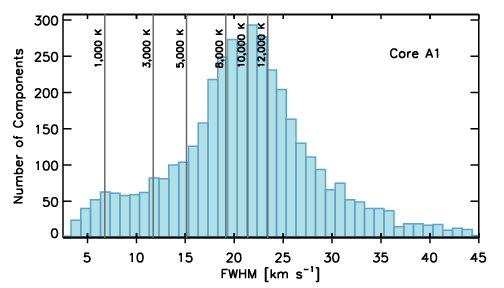}\includegraphics[trim=33 25 15 10,clip,scale=0.35,angle=0]{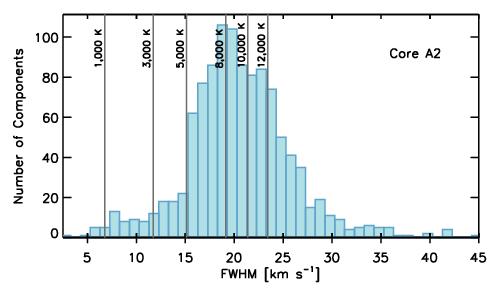} \\
\includegraphics[trim=10 25 15 5,clip,scale=0.35,angle=0]{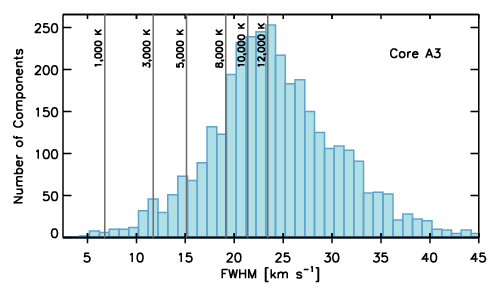}\includegraphics[trim=33 25 15 10,clip,scale=0.35,angle=0]{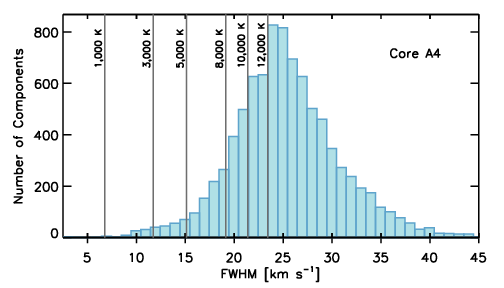}\includegraphics[trim=33 25 15 10,clip,scale=0.35,angle=0]{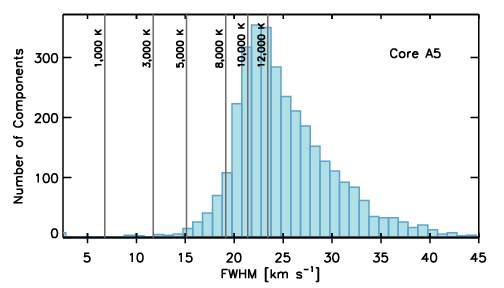} \\
\includegraphics[trim=10 0 15 10,clip,scale=0.35,angle=0]{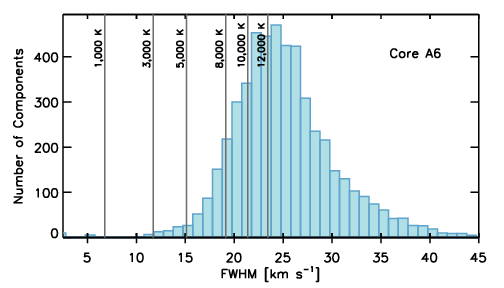}\includegraphics[trim=33 0 15 10,clip,scale=0.35,angle=0]{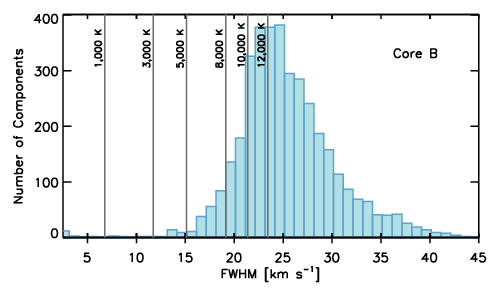}\includegraphics[trim=33 0 15 10,clip,scale=0.35,angle=0]{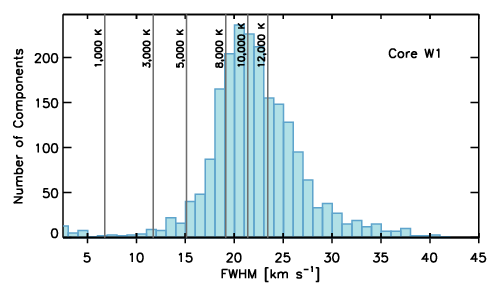} \\ \includegraphics[trim=33 0 15 10,clip,scale=0.35,angle=0]{Figures/fig10i.png} 
\end{center}
\figcaption{Histogram distributions of the ${\rm FWHM}$ for each of the major \hi\ core regions (see Figures~\ref{figure:decomp_region_A0}--\ref{figure:decomp_region_B}) and the two high-latitude wing regions (see Figures~\ref{figure:decomp_region_W1}--\ref{figure:decomp_region_W2}). These line widths were through determined by decomposing the \hi\ spectra into Gaussian components (see Section~\ref{section:gauss_decoms}). We additionally mark the gas temperature that these widths would correspond to for a pure thermal broadening scenario, where the 8,000$\lesssim T\lesssim$12,000~K lines is where \ha\ emission peaks. 
\label{figure:histograms}}
\end{figure*}

\begin{figure*}
\begin{center}
 \includegraphics[trim=0 52 0 0,clip,scale=0.40,angle=0]{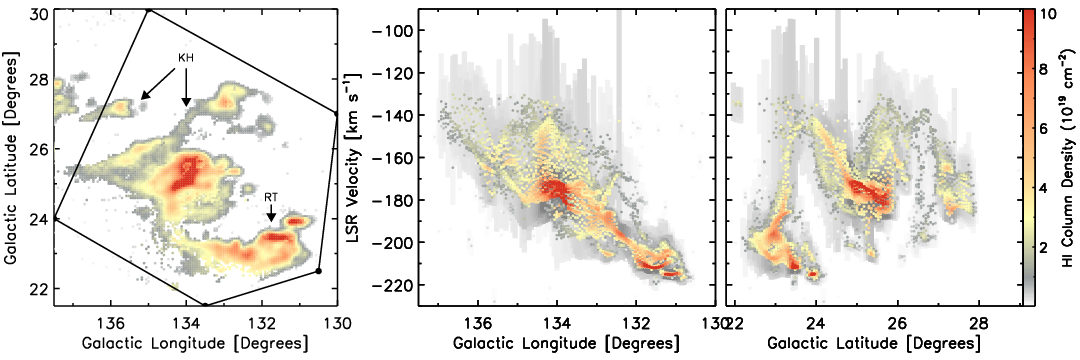} \\
 \includegraphics[trim=0 7 0 0,clip,scale=0.40,angle=0]{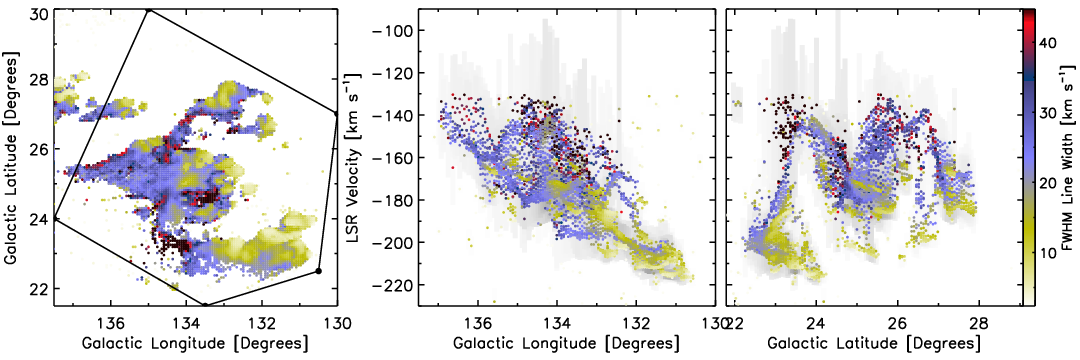} 
\end{center}
\figcaption{Gas distribution of the core~A0 region over the $-230\le\rm v_{LSR}\le-130~\kms$ velocity range with the  Rayleigh-Taylor (RT) globules, Kelvin-Helmholtz structures, and ram-pressure stripping (RPS) features labeled in the Top-Left panel. The colors in the top row of panels represent the H\textsc{~i} column density and the colors on the bottom panels represent the FWHM line width. All of the values were determined through Gaussian decompositions (see Section~\ref{section:gauss_decoms}). Left Columns: Galactic Latitude and Longitude position-position map. Middle Columns: Galactic longitude position-velocity map. Right Columns: Galactic latitude position-velocity map. The panels in the Left Column only display the $\rm N_{\rm H\textsc{~i}}$ and ${\rm FWHM}$ for the component with the highest \hi\ column density along each sightline, though all components are included in the Middle and Right Column position-velocity maps. Only the emission contained within the spatial region marked black boundary in the Left panel is the included in the Middle and Right position-velocity maps. The faint gray-scale vertical bars the in the background of the position-velocity map represent the kinematic extent of the emission and span from $\rm v_{\rm extent}=\rm v_{\rm center}\pm \Delta v_{\rm width}$. 
\label{figure:decomp_region_A0}}
\end{figure*}

\begin{figure*}[t]
\begin{center}
 \includegraphics[trim=0 52 0 0,clip,scale=0.40,angle=0]{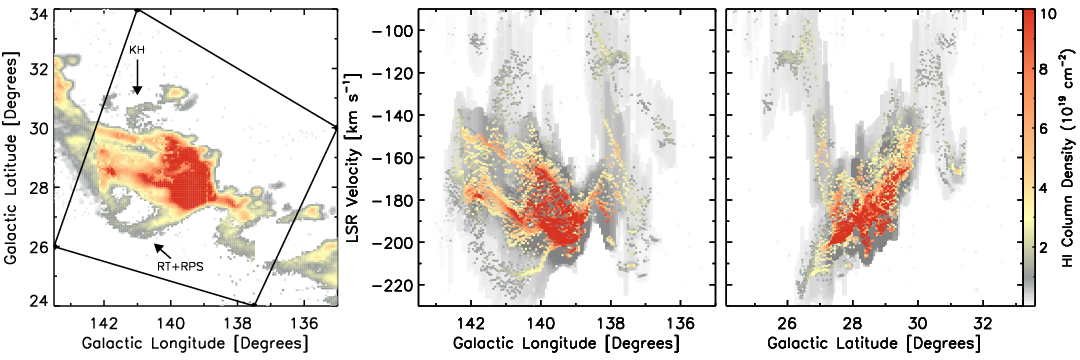} \\
 \includegraphics[trim=0 7 0 0,clip,scale=0.40,angle=0]{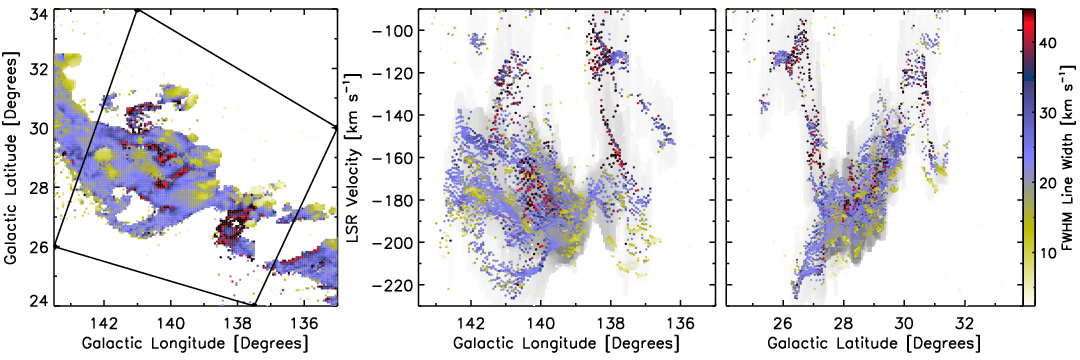} 
\end{center}
\figcaption{Same as Figure~\ref{figure:decomp_region_A0}, but for the core~AI region over the $-230\le\rm v_{LSR}\le-90~\kms$ velocity range. 
\label{figure:decomp_region_A1}}
\end{figure*}

\section{Data Reduction}\label{section:reduction}

For this project, we used a reduction and calibration procedure that is very similar to the one used by \citet{2010ApJ...723.1618N} with only a few small modifications.
We calibrated antenna temperature ($T_A$) of our frequency-switched H\textsc{~i} 21-cm dataset comparing the flux of our target with the flux of a well characterized objects,\footnote{We performed this calibration with the GETFS program in the GBTIDL (http://gbtidl.nrao.edu/) software.} using the relationship: 
\begin{equation}\label{eq:Ta}
T_{\rm A,~calib}(\nu) = T_{\rm sys}^{\rm ref}(\nu) \times \frac{{F_{\rm sig}(\nu)-F_{ref}(\nu)}}{F_{\rm ref}(\nu)}
\end{equation}
Here, the $T_{\rm sys}^{\rm ref}(\nu)$ is the system temperature that we determined from the reference spectrum, $F_{\rm ref}(\nu)$ is the reference flux of a calibration target, and $F_{\rm sig}(\nu)$ is the flux of the on-target signal. We assumed that brightness temperature ($T_{\rm B}$) is roughly equal to the antenna temperature (i.e., $T_{\rm B}\approx T_A$). Our calibration objects included standard S7 and S8 reference targets located at $(l,~b)=(207\fdg00,~-1\fdg00)$ and $(207\fdg00,~-15\fdg00)$ \citep{1973A&AS....8..505W}. We additionally observed the center of the core~AVI at $(160\fdg23,~43\fdg04)$ during each observational run, enabling us to measure the H\textsc{~i} emission along this sightline very accurately and to use it as a substitute flux calibration target whenever both S7 and S8 were below the horizon.

Once calibrated, we binned the spectra to a $\Delta \rm v_{\rm bin}=0.8~\kms$ velocity spacing to decrease small scale fluctuations in the signal and further smoothed the spectra with a Gaussian kernel and then removed the continuum level. We fit the baseline for each integration and polarization separately after masking out the Galactic emission between $-100\lesssim\rm v_{LSR}\lesssim +100~\kms$ and emission lines above 2-sigma. For the XX polarization, we fit a 5th-order polynomial to the continuum. We used the same procedure for the YY polarization with the addition of a sinusoidal component in the fit to remove a standing wave that has a period of $\nu\approx1.6~\rm MHz$ in the GBT spectra (see \citealt{2010ApJ...723.1618N}). Next, we centered the baseline at $T_{\rm B}(\rm v)=0~{\rm mK}~\kms$ by subtracting the median emission--free spectral height of all spectra in an observing session from each polarization. Finally, we averaged the two polarizations together to produce reduced spectra. The resultant spectra has a typical root-mean-square (RMS) noise that is $T_{\rm B}\approx75~\rm mK$ per $0.8~\kms$ channel. This corresponds to a spectral noise sensitivity of $\log({N_{\rm H\textsc{~i},\,1\sigma}}/\cm^{-2})=17.7$ for a line with a half-maximum of ${\rm FWHM}=20~\kms$ (see Figures~\ref{figure:noise_map} and~\ref{figure:residuals}), using 
\begin{equation}
\frac{N_{\rm H\textsc{~i},\,1\sigma}}{\cm^{-2}}=1.822\times10^{18}\left(\frac{T_{\rm B,~1\sigma}}{\rm K}\right)\sqrt{\frac{{\rm FWHM}}{\kms} \frac{\Delta \rm v_{\rm channel}}{\kms}}
\end{equation}
to convert between $T_{\rm B}$ and $N_{\rm H\textsc{~i}}$ noise sensitivity under the assumption that the spectral noise is Gaussian in nature \citep{2014PhDT...Wolfe}. Therefore, our 3-sigma detection limit is $\log({N_{\rm H\textsc{~i},~3\sigma}}/\cm^{-2})=18.2$ for a line with a ${\rm FWHM}=20~\kms$ width. We determined the $\rm N_{\rm H\textsc{~i}}$ of our lines from the $\rm T_{\rm B}$ under the assumption that emitting gas is optically thin to self absorption:
\begin{equation}
\frac{\rm N_{\rm H\textsc{~i}}}{\cm^{-2}}=1.822\times10^{18}\int \left(\frac{\rm T_{\rm B}(\rm v)}{\rm K}\right)\left(\frac{\rm dv}{\kms}\right)
\end{equation}

For the archival datasets, we use the already reduced and calibrated observations that were shared with us by those program leaders. The reduction and calibration procedures outlined above for our new GBT observations from the GBT13B-068 (PI Barger) program were the same procedures used to reduce the GBT10A-003 (PI Verschuur) dataset, which they outline in their study that explored the physical conditions of Complex~A's core~A0 \citep{2013ApJ...766..113V}. The calibration and continuum level removal techniques in for the GBT07A-104 (PI Martin) observations are described in \citet{2011A&A...536A..81B} and \citet{2015ApJ...809..153M}.

The observation and reduction procedures used on the GBT09A-046 (PI Chynoweth) dataset differs substantially from the other datasets. These observations were taken in position switching mode instead of frequency-switching mode that was used for all other datasets in our survey. For each observing session, \citet{2011AJ....141....9C} used a reference spectrum positioned at the edge of their observing grid as a reference signal as their flux calibrator. Through this calibration scheme, most of the Milky Way zero-velocity is removed, but the Complex~A features remain essentially intact. The spectra was then binned to $\Delta\rm v_{bin}=5.2~\kms$. 

We combined and resampled all of the new and archival datasets---except the GBT09A-046 (PI Chynoweth) dataset---on a large, uniform grid in Galactic coordinates with $\Delta\theta=3\farcm5$ spatial steps and $\Delta \rm v=0.8~\kms$ velocity bins. Because the velocity sampling and angular resolution of the GBT09A-046 (PI Chynoweth) dataset differ the most from the other datasets, we only used those observations when other data were not available. In \autoref{figure:noise_map} we shows the 1$\sigma$ sensitivity map of our all observations.

\section{Gaussian Decompositions}\label{section:gauss_decoms}

\begin{figure*}[t]
\begin{center}
 \includegraphics[trim=0 52 0 0,clip,scale=0.40,angle=0]{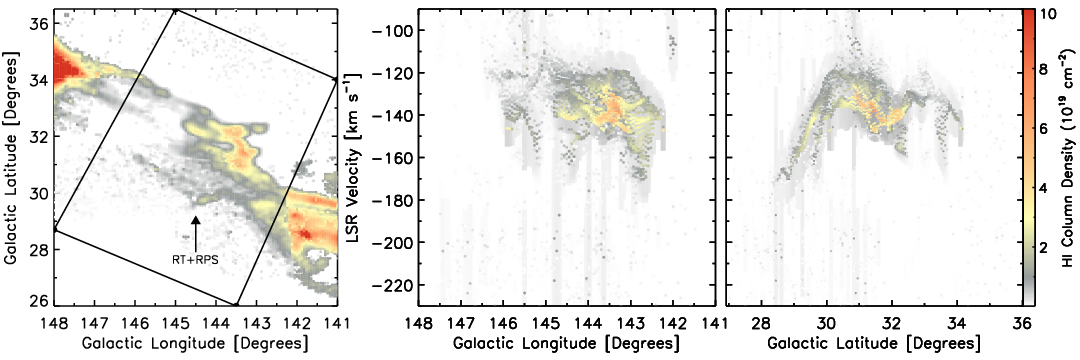} \\
 \includegraphics[trim=0 7 0 0,clip,scale=0.40,angle=0]{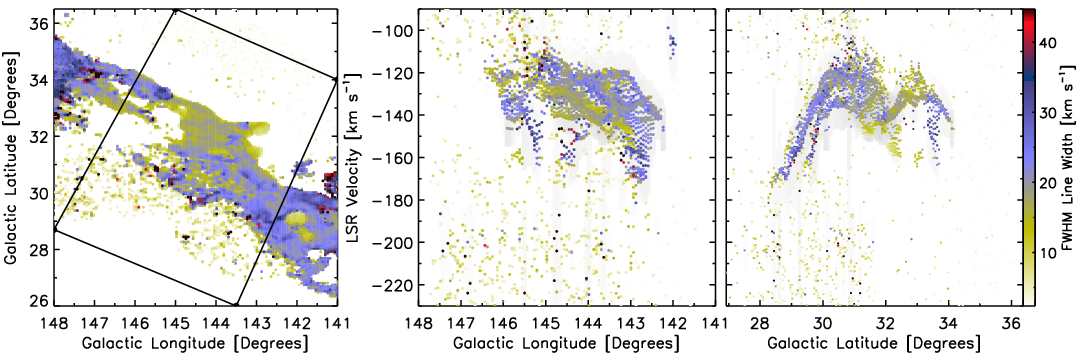}
\end{center}
\figcaption{Same as Figure~\ref{figure:decomp_region_A1}, but for the core~AII region. 
\label{figure:decomp_region_A2}}
\end{figure*}

\begin{figure*}[t]
\begin{center}
 \includegraphics[trim=0 52 0 0,clip,scale=0.40,angle=0]{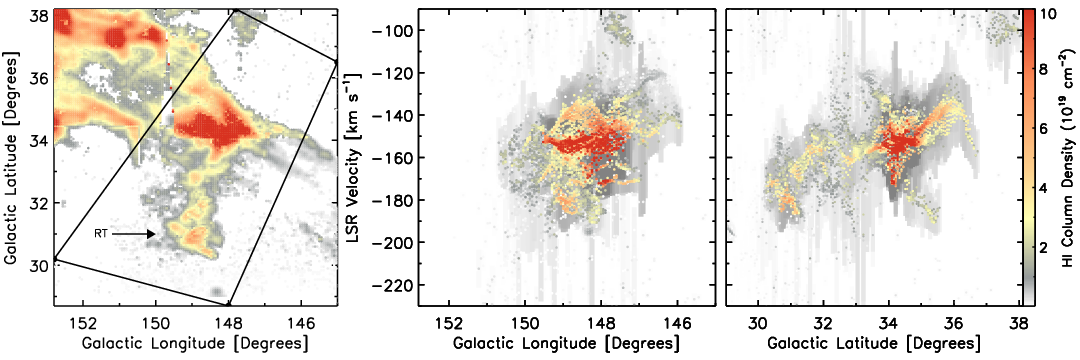} \\
 \includegraphics[trim=0 7 0 0,clip,scale=0.40,angle=0]{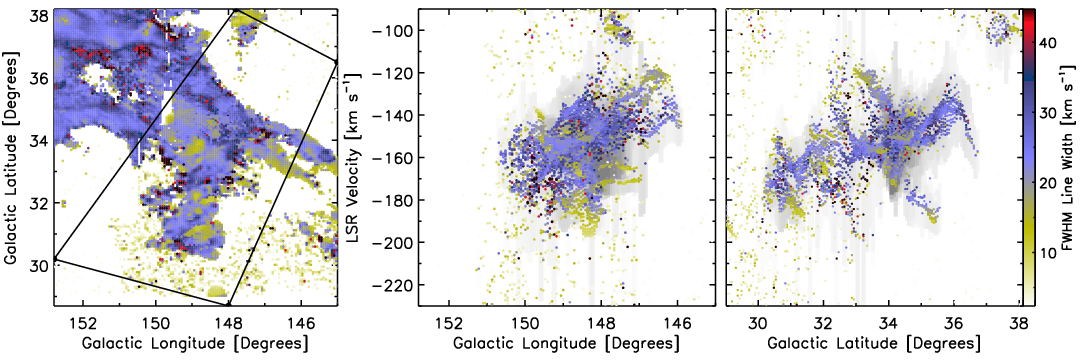}
\end{center}
\figcaption{Same as Figure~\ref{figure:decomp_region_A1}, but for the core~AIII region.
\label{figure:decomp_region_A3}}
\end{figure*}

\begin{figure*}[t]
\begin{center}
 \includegraphics[trim=0 52 0 0,clip,scale=0.40,angle=0]{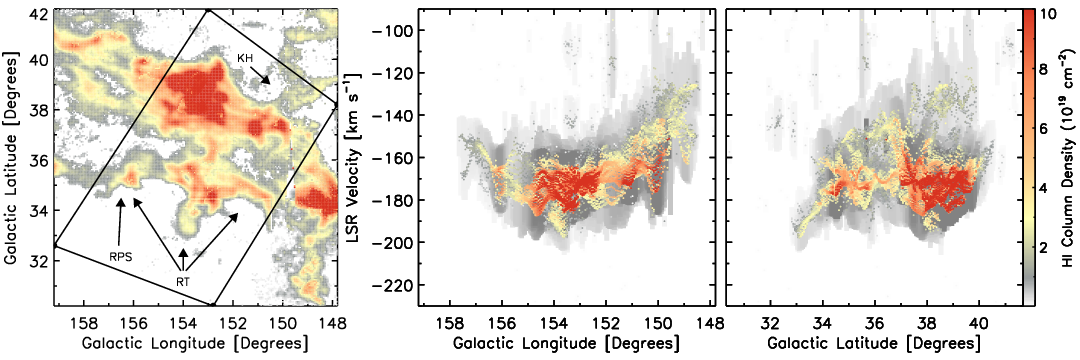} \\
 \includegraphics[trim=0 7 0 0,clip,scale=0.40,angle=0]{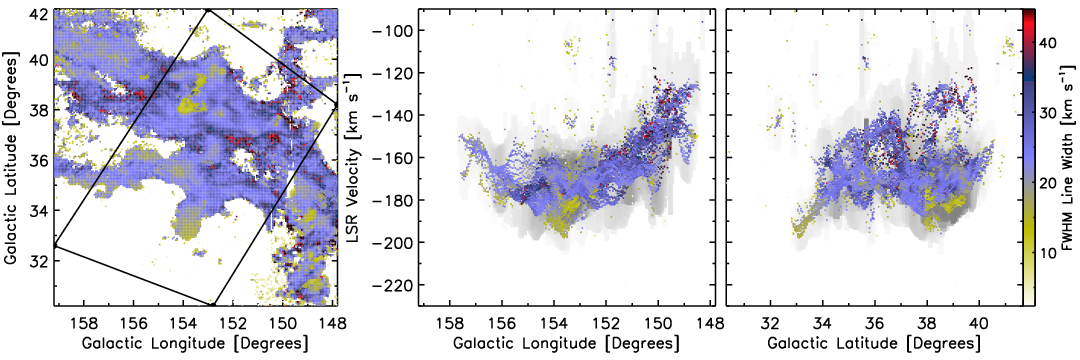}
\end{center}
\figcaption{Same as Figure~\ref{figure:decomp_region_A1}, but for the core~AIV region.
\label{figure:decomp_region_A4}}
\end{figure*}

\begin{figure*}[t]
\begin{center}
 \includegraphics[trim=0 52 0 0,clip,scale=0.40,angle=0]{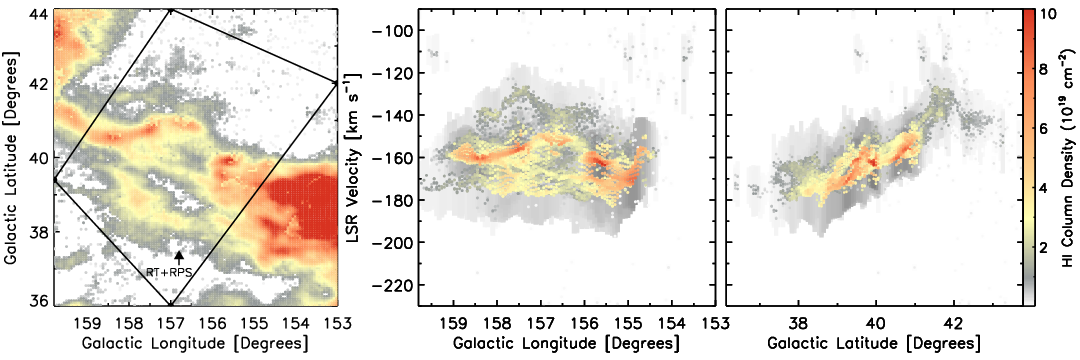} \\
 \includegraphics[trim=0 7 0 0,clip,scale=0.40,angle=0]{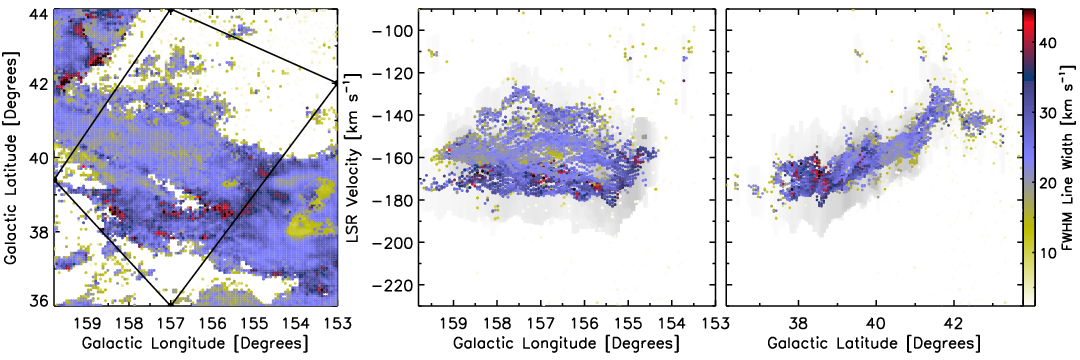} 
\end{center}
\figcaption{Same as Figure~\ref{figure:decomp_region_A1}, but for the gas distribution of the core~AV region.
\label{figure:decomp_region_A5}}
\end{figure*}

\begin{figure*}[t]
\begin{center}
 \includegraphics[trim=0 52 0 0,clip,scale=0.40,angle=0]{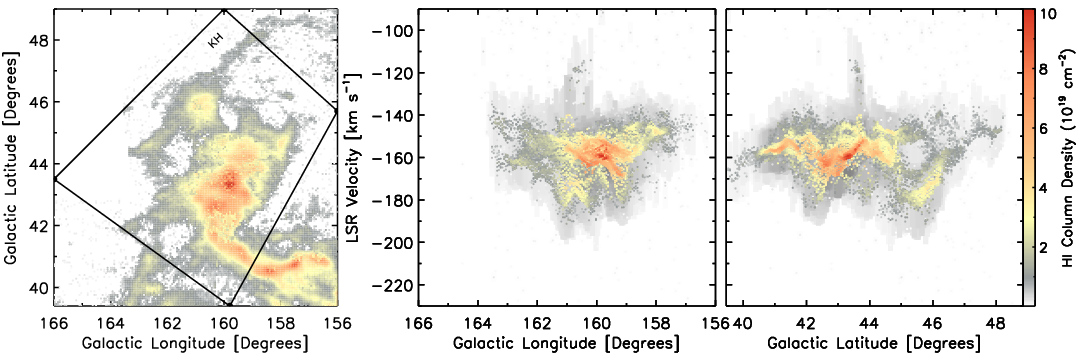} \\
 \includegraphics[trim=0 7 0 0,clip,scale=0.40,angle=0]{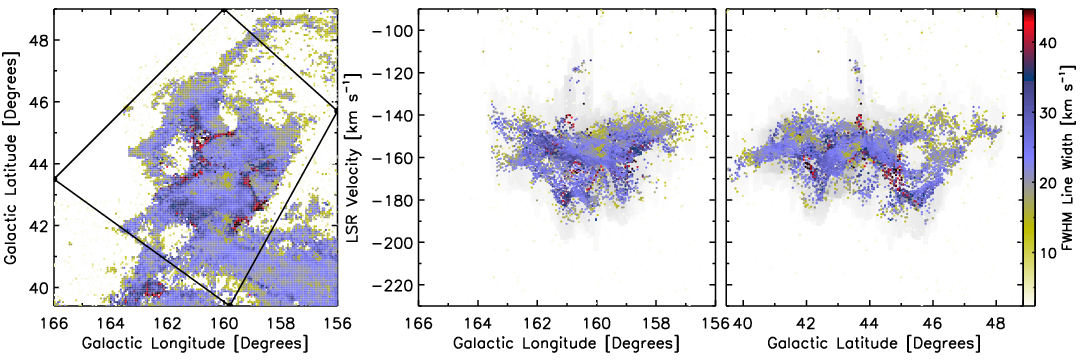}
\end{center}
\figcaption{Same as Figure~\ref{figure:decomp_region_A1}, but for the gas distribution of the core~AVI.
\label{figure:decomp_region_A6}}
\end{figure*}

\begin{figure*}[t]
\begin{center}
 \includegraphics[trim=0 52 0 0,clip,scale=0.40,angle=0]{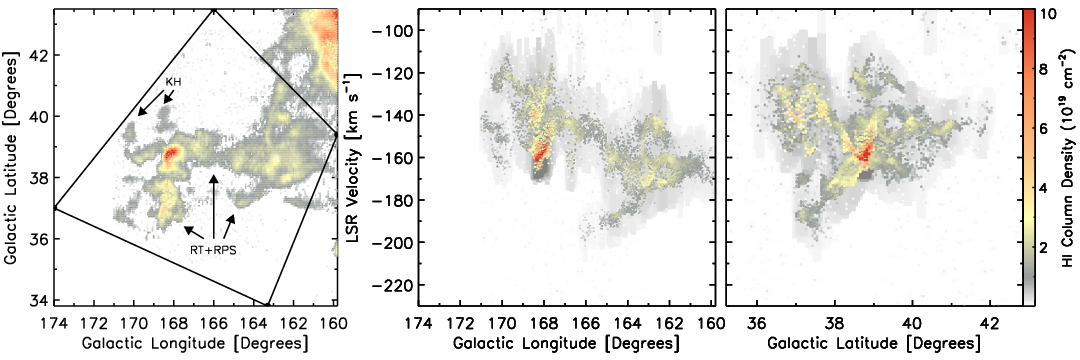} \\
 \includegraphics[trim=0 7 0 0,clip,scale=0.40,angle=0]{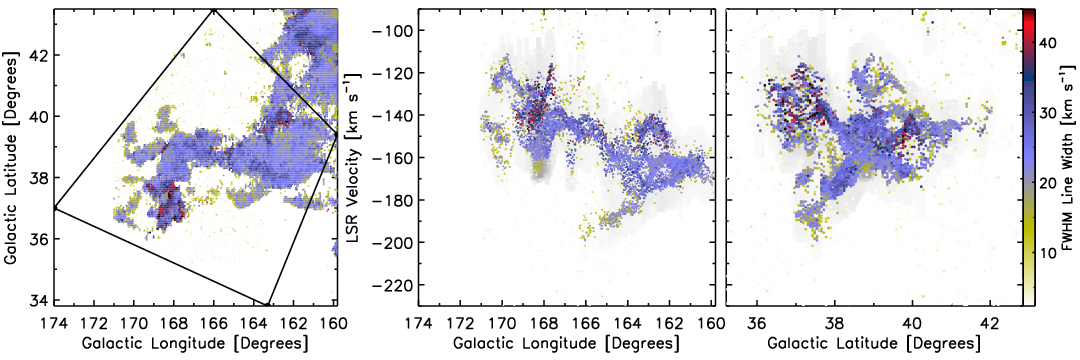}
\end{center}
\figcaption{Same as Figure~\ref{figure:decomp_region_A1}, but for the core~B region.
\label{figure:decomp_region_B}}
\end{figure*}

\begin{figure*}[t]
\begin{center}
 \includegraphics[trim=0 52 0 0,clip,scale=0.40,angle=0]{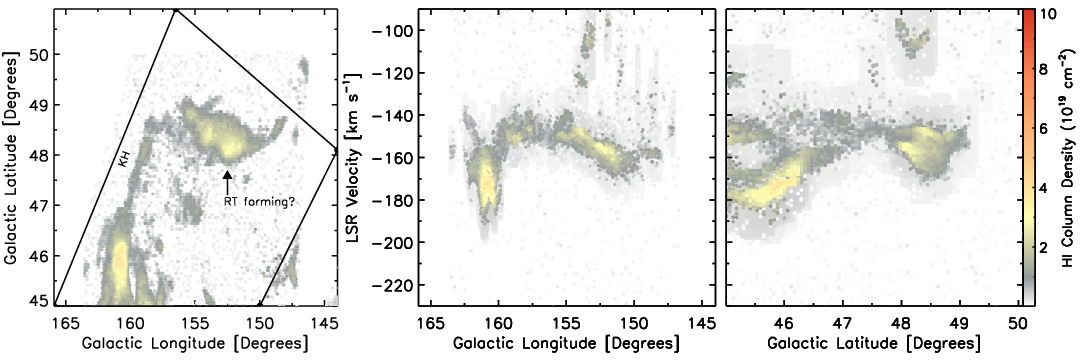} \\
 \includegraphics[trim=0 7 0 0,clip,scale=0.40,angle=0]{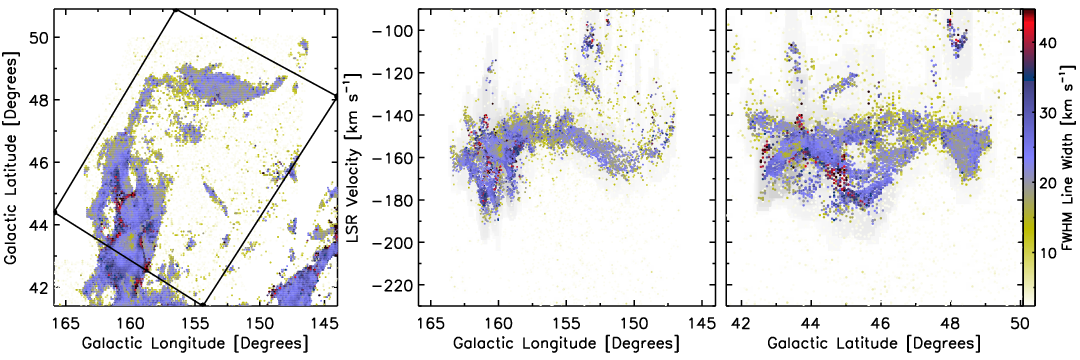}
\end{center}
\figcaption{Same as Figure~\ref{figure:decomp_region_A1}, but for wing~1 (W1)---a the high-latitude cloud fragment that lies off the core~AVI region at $(l,~b)\approx(154\fdg4,~48\fdg4)$.  
\label{figure:decomp_region_W1}}
\end{figure*}

\begin{figure*}[t]
\begin{center}
 \includegraphics[trim=0 52 0 0,clip,scale=0.40,angle=0]{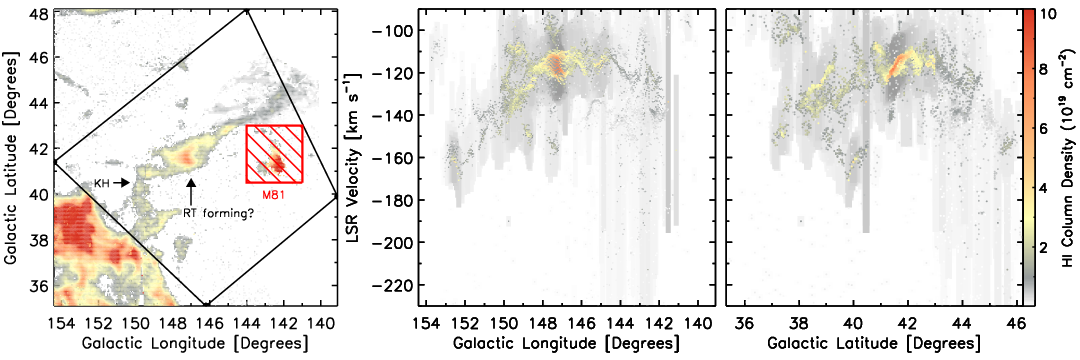} \\
 \includegraphics[trim=0 7 0 0,clip,scale=0.40,angle=0]{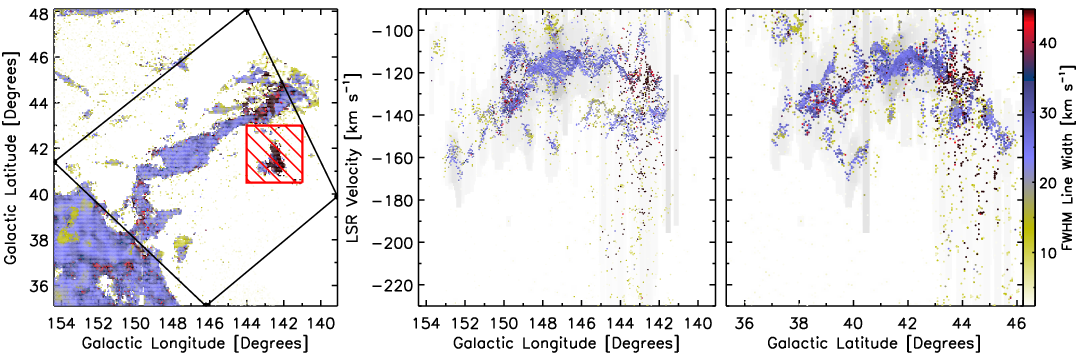}
\end{center}
\figcaption{Same as Figure~\ref{figure:decomp_region_A1}, but for wing~2 (W2)---a high-latitude cloud fragment that lies off the core~AIV region at $(l,~b)\approx(147\fdg6,~41\fdg6)$. Just offset the tip of this wing, there is a high column density ($\log{\left(N_{\rm H\textsc{~i}}/\cm^2\right)}\approx20$) cloudlet at $(142\arcdeg,~41\arcdeg)$ that is associated with M81 galaxy, which is enclosed within a red rectangle in the left-hand panels. This M81 emission has been removed from the center and right-hand panels. 
\label{figure:decomp_region_W2}}
\end{figure*}

We determined the component structure of the H\textsc{~i}~21-cm emission of Complex~A by modeling its emission lines with Gaussian profiles. We characterized the area, center position, and width for each fitted emission line. We determined the number Gaussians to model the emission by minimizing the reduced chi-squared ($\widetilde{\chi}^2$) with the MPFIT\footnote{The IDL MPFIT routines are available at http://purl.com/net/mpfit.} IDL routine \citep{2009ASPC..411..251M}. We selected the fit that used the smallest number of Gaussians to achieve a reduced-chi squared that was within 0.25 of the best fit to avoid arbitrarily adding more and more Gaussian profiles that are not of physical significance to better match the \hi\ emission. 

Our initial guesses for the Gaussian fit parameters determined by iteratively searching for peaks in each spectrum. We identified peaks as locations with the highest $N_{\rm H\textsc{~i}}\left(\rm v\right)$ emission above the $1$~standard deviation noise level of a smoothed spectrum. We then masked out all regions with $\rm v_{peak}\pm25~\kms$ and repeated this search for emission-lines in the unmasked region. We then fit the spectrum using for lines at velocity positions with $\Delta\rm v=10~\kms$ for the initial guess of the line width. We checked the quality of each fit by searching the residuals for any remaining unfitted emission lines, where the residuals are the spectral signatures remaining after the fit is subtracted from the H\textsc{~i} emission spectrum:
\begin{equation}\label{eq:residuals}
{\rm residuals}({\rm v}) = N_{\rm H\textsc{~i},~obs}({\rm v})-N_{\rm H\textsc{~i},~fit}({\rm v})
\end{equation}

We only report results for fitted line profiles that have a signal-to-noise ratio ($\rm S/N$) greater than~$2$ and that are above the 3-sigma detection limit of our survey at $\log({N_{\rm H\textsc{~i},~3\sigma}}/\cm^{-2})\ge17.7$. We defined the area of the signal in the $\rm S/N$ to as the area of the fitted Gaussian-line profile and the area of the noise to be equivalent to the area of a rectangle that has a height equal to the standard deviation of the continuum and a width equal to the FWHM of the fitted Gaussian-line profile. This step was done to remove unrealistic fits that characterized spikes in the noise and to ensure that the faint emission that is associated with the diffuse outer envelope of Complex~A was kept. 

In Figure~\ref{figure:spectra}, we illustrate representative Gaussian decompositions of three sightlines along cores~AIII, AVI, and~B and the corresponding residuals of those fits. For each of these sightlines, there is a brighter component that is associated with the H\textsc{~i} core. Additionally, we often find fainter and wider components, which trace the diffuse gas in the outer envelope of this HVC. We include a movie that rotates from through a 3~dimensional position-position-velocity map of Complex~A that shows all of the LSR line center and $N_{H\textsc{~i}}$ solutions to our Gaussian decomposition fits in Figure~\ref{figure:3d_decompositions}. We additionally provide a vertically stacked spectral image of the residuals of all $\sim1.2\times10^5$ sightlines explored in this study in Figure~\ref{figure:residuals}, which are typically $\log{\left(\rm residuals(v)/[\cm^{-2}~\kms]\right)<17.5}$ per velocity bin. 

\section{Complex A Coordinate System}\label{section:complexa_coords}

Because Complex~A has a long, filamentary structure it is useful to have a coordinate system where the equator lies along the great circle of this extended \hi structure.  We define a ``Complex~A'' coordinate system with a pole at $(l,~b)$=(202\degr,$-$40\degr) and the origin of the longitude axis (the ascending node) defined such that the center of the A0 core at $(l,~b)=(133\fdg9,~+25\fdg1)$ corresponds to $(l_{\rm CA},\,b_{\rm CA})=(0\arcdeg,\,0\arcdeg$). As in the Magellanic \citep{2001ApJS..136..463W} and Magellanic Stream coordinate systems \citep{2008ApJ...679..432N}, $l_{\rm CA}$ decreases along Complex~A towards higher Galactic latitudes.  Figure~\ref{figure:alonalat_coldens} shows the column density of Complex~A and Figure~\ref{figure:alonvlsr} shows the position--velocity diagram in this new coordinate system.

\section{Neutral Gas Morphology and kinematics}\label{section:kinematics_morphology} 

\subsection{Global Properties}

Complex~A is an elongated stream with multiple dense H\textsc{~i} cores along its length (see Figures~\ref{figure:all_regions} and~\ref{figure:alonalat_coldens}). These cores tend to be more compressed on the lower Galactic latitude and longitude side (or higher $l_{\rm CA}$ side) and more diffuse and elongated on the opposite side, indicating that core~A0 represents the leading end of this stream and core~AVI represents the trailing end (see more discussion below). This stream is wider at its trailing end  (see Figures~\ref{figure:all_regions} and~\ref{figure:alonalat_coldens}). Because the leading edge of Complex~A is much closer ($d_\odot\approx6~\kpc$) than the tailing gas ($d_\odot\approx10~\kpc$; see \citealt{2012ApJ...761..145B}), this means that the wider angular extent of cores~AVI and~B corresponds to a \textit{much} larger physical width at $\Delta\theta_{\rm core\,A0}\approx0.10~{\kpc}/{\rm degree}$ vs $\Delta\theta_{\rm core\,AVI}\approx0.17~{\kpc}/{\rm degree}$. However, the relatively inline A0--AVI cores suggests that they are part of the main body of Complex~A and that core~B represents material that fractured off this gas stream.

There is a relatively coherent Galactic standard of rest (GSR) velocity gradient along the length of Complex~A, where its leading gas traveling slower relative to the Milky Way than its trailing gas (see lower right-hand panel in Figure~\ref{figure:all_regions}); this indicates that that Complex~A is \textit{deccelorating}. The GSR velocity gradient along its $\Delta\theta \approx 33\arcdeg$ (or $\Delta L=5.7~\kpc$) body relative to its leading edge is $\Delta\rm v_{GSR}/\Delta\theta=4.2~\kms/{\rm degree}$ (or $\Delta\rm v_{GSR}/\Delta L=25~\kms/{\kpc}$). Assuming an average velocity of $\langle \rm v_{GSR}\rangle\approx -70~\kms$, it has taken Complex~A $\Delta t=80~{\rm Myrs}$ to travel the length of its body, corresponding to decceleration of $\langle a\rangle_{\rm GSR}=55~{\rm km}/{\rm yr}^2$. At a constant acceleration, this complex will cross the Galactic plane at $b=0\arcdeg$ in $\Delta t\lesssim70~{\rm Myrs}$; this is an upper limit as this time should decrease due to an increasing gravitational pull as the Complex~A approaches the Milky Way's.

We also find that there is a graduate increase in the width of the \hi\ line toward the trailing end of Complex~A. In Figure~\ref{figure:clon_temp}, we have plotted the median line widths for each of the major core regions and the two high-latitude wings as a function of Complex~A longitude. In the core~A0 region, the median FWHM line width is roughly ${\rm FWHM}\approx19~\kms$ at $l_{\rm CA}\approx0\arcdeg$, but grows to ${\rm FWHM} \approx25~\kms$ at $l_{\rm CA}\approx-28\arcdeg$ for cores~A6 and~B. We have included the histogram distributions of the line widths for all fitted components in each of these regions in Figure~\ref{figure:histograms}. In general, the histogram distributions for each core region are relatively well behaved with an easy to identify peak in the number of components at a particular line width, except for the core~A0. For this leading core, the line widths peak between $10\lesssim{\rm FWHM}\lesssim23~\kms$ and include a much larger distribution of narrow lines than any other core region. These narrow lines suggest that this core is cooling rapidly, presumably because this low metallicity core is mixing with the higher metallicity gas near the Milky Way's disk. 

Assuming that the \hi\ emission lines are only broadened by thermal broadening, the increasing median line widths along the length of Complex~A would correspond to a rise in the hydrogen gas temperature by roughly 4,400~K from $T_{\rm H\textsc{~i},\,median}=$~8,700~K along the leading edge of the complex to 13,100~K along its trailing edge (see Figure~\ref{figure:clon_temp}). Overall, this is relatively inline with the  typical gas temperature that \citet{2012ApJ...761..145B} found for the warm ionized phase of this complex at $T_{\rm H\alpha}=12,600~\rm K$ in the direction of the \hi\ cores, where their WHAM \ha\ observations were resolved at $\Delta\theta=1\arcdeg$ and have an angular area that is larger than the GBT \hi\ observations by a factor of $A_{\theta,\,\rm WHAM}\approx43 A_{\theta,\,\rm GBT}$.
 However, the elevated line widths on the trailing end of Complex~A could also signify that this gas is experiencing an increase in non-thermal motions. If that is the case, then the emission lines associated with the trailing end of Complex~A would be additionally non-thermally broadened by $\rm FWHM_{non-thermal}=14.1~\kms$, assuming a thermal broadening of $\rm FWHM_{thermal}\approx20~\kms$. This significant contribution to the non-thermal broadening of the line width would indicate that the trailing gas is being more disrupted. 

The higher \hi\ column density cores in this stream are connected by lower column density gas. Many of the \hi\ cores are compressed in the direction of their motion, including cores~A0, AI, AII, and AIII. These  cores also tend to be moving towards the Galaxy \textit{faster} (i.e., larger negative LSR velocities) than the lower column density gas that surrounds and connects them. This global trend is especially apparent in movie found in Figure~\ref{figure:3d_decompositions}, which rotates Complex~A through 3~dimensional position-position-velocity space, and in Figure~\ref{figure:alonvlsr}. We additional provide two sets of zoomed in position-position and position-velocity maps that are scaled by the \hi\ column density and FWHM line width of each core region in Figures~\ref{figure:decomp_region_A0}--\ref{figure:decomp_region_W2}. This global morphology is characteristic of ram-pressure stripping in which the surrounding coronal gas and incident Galactic photons act as a headwind that compresses the leading gas and strips the outer layers of this stream to form a lagging diffuse tail that travels in the anti-direction of motion. 

The fragmented morphology of Complex~A could be a result of thermal cooling instabilities or a slow ``shock cascade.'' Cooling instabilities often arise as a result of density inhomogeneities in which the high density gas cools more efficiently. As the complex descends toward the disk, it will sweeps up coronal gas, compressing the leading gas \citep{2009ApJ...700L...1K}. Further, gas that the complex sweeps up gas near the Milky Way's disk will have a higher metallicity (${Z}_{\rm CA}=0.1~{Z}_\odot$: \citealt{1994A&A...282..709K, 1995A&A...302..364S,  1999Natur.400..138V, 2001ApJS..136..463W, 2012ApJ...761..145B} and will promote cooling. Fragmentation is expected to occur once the stream sweeps up roughly its own mass in ambient material \citep{2004ApJ...615..586M}, indicating that Complex~A has already accreted a substantial material during its journey. Unfortunately, the sparseness of metallicity measurements along the length of the complex means that the level of metal mixing and accretion cannot currently be constrained. In a ``shock cascade'' scenario, leading material that is stripped via ram-pressure will be slowed by non-conservative forces and can then collide with down stream gas \citep{2007ApJ...670L.109B, 2015ApJ...813...94T}. This shock cascade can disrupt and and fragment the down stream gas. The rapidly varying line widths with position and column density, between $10\lesssim {\rm FWHM}\lesssim 40~\kms$  (see Figures~\ref{figure:decomp_region_A0}--\ref{figure:decomp_region_W2}) are a strongly indicator that the low and high column density H\textsc{~i} gas is either not in thermal-dynamical equilibrium or that the low column density gas is experiencing more severe non-thermal motions.

Interestingly, the gas in the core~AII region has a much lower \hi\ column density than the gas in the adjacent cores that connect to it. Further, this relatively wispy core region is moving much slower toward the Galactic disk than cores~AI and~AIII  at $\Delta\rm v_{LSR}\approx80~\kms$ offset from core~AI and $\Delta\rm v_{LSR}\approx30~\kms$ offset from core~AIII (see Figure~\ref{figure:alonvlsr}), indicating that it is much more influenced by coronal-gas interactions. However, although core~AII is morphologically much more disrupted, its higher column density gas still has a narrow line profile ($10\lesssim \rm {FWHM} \lesssim 20~\kms$; Figure~\ref{figure:decomp_region_A2}). This suggests that the gas in the core~AII region is still able to remain relatively cool and compact. Using mapped \ha\ observations, \citet{2012ApJ...761..145B} found that roughly half of Complex A is ionized. This warmer and lower density phase acts as skin that shields the H\textsc{~i} cores from direct interactions with the surrounding coronal gas. The gas in core~AII could also be ``drafting'' the leading gas in core~AI, such that it is not experiencing a direct headwind, though we do not know the locations of these cores in 6-dimensional position and velocity space and therefore cannot tell how well aligned core~AII is behind core~AI.

Numerous H\textsc{~i} structures protrude or are fractured off Complex~A's main body, which is an indication that its gas is subject to hydrodynamic instabilities. As these offset structures still have a relatively cohesive structure in H\textsc{~i}, they were likely recently stripped off the complex. This displaced gas is now more exposed to the incident ionizing radiation and the surrounding coronal gas as this material now has a larger surface area and is no longer ``drafting'' behind the cloud. This increased exposure will cause them to be heated and ionized quicker, which will lead to them rapidly evaporating \citep{2002A&A...391..713K}. We identify these offset structures and discuss how Rayleigh-Taylor and Kelvin-Helmholtz instabilities are working with ram-pressure stripping producing these structures in the following subsections.

\subsection{Rayleigh-Taylor Instability Structures}\label{subsection:RT}

Complex~A is surrounded by hot coronal gas, which means that it is essentially resting on top of a lower density medium while being influenced by the Milky Way's gravitational field. This is an unstable configuration that can drive buoyancy related disturbances known as Rayleigh-Taylor instabilities. If these instabilities are strong enough, they can generate globules and spikes (``finger'' like structures) that drip through the warmer coronal medium that lies below Complex~A toward the center of the Milky Way's gravitational field. These globules will therefore form on the lower Galactic latitude edge of this complex (see Figure~\ref{figure:alonalat_coldens}). Further, the morphology will include a compressed edge that forms when globules push through the lower density coronal gas below it. 

There is an \hi\ arch that hangs off core~A0 at $(l,~b)\approx(132\arcdeg,~23\arcdeg)$ by a thin filament (see Figure~\ref{figure:decomp_region_A0}). The high-latitude portion of this arch is more compressed, indicating that it was the material that initially pushed through the coronal gas when the globular began its departure from core~A0. This gas arches in Complex~A's direction of motion, which is unusual as ram-pressure stripping should have pushed this material in the other direction. Instead, as core~A0 is only $z\approx2.7~\kpc$ above the Galactic plane \citep{2012ApJ...761..145B}, this arched morphology in the direction of motion was likely created when the globular interacted with the denser gaseous medium near the Milky Way's disk. The thin connecting filament further indicates that this globular will soon fracture off core~A0. 

On the lower latitude edge of core~AI, there is an \hi\ structure that looks like a skewed ``loop'' at $(l,~b)\approx(141\arcdeg,~26\fdg5)$ (see Figure~\ref{figure:decomp_region_A1}). This material connects to core~AI at $(140\arcdeg,~27\arcdeg)$ and extends in the anti-direction of Complex~A's motion and then curves back up toward core~AI. This loop appears to represent a Rayleigh-Taylor spike that was elongated and pushed backward due to ram-pressure stripping by the surrounding coronal gas. A Rayleigh-Taylor spike also lies on the lower latitude side of core~AIII at $(1,~b)\approx(149\arcdeg,~31\arcdeg)$ (see Figure~\ref{figure:decomp_region_A3}). However, this relatively shorter and wider structure projects downward in Galactic latitude and does not curve backward, indicating that this gas only recently ``dripped'' off core~AIII.

There is a mini stream that branches off core~AIV's lower latitude edge and points in the anti-direction of Complex~A's motion (see Figure~\ref{figure:decomp_region_A4}). This mini stream represents a complex Rayleigh-Taylor instability structure that is strongly influenced by ram-pressure stripping. The gas that is positioned directly under core~AIV has only recently ``dripped'' off core~A4. Below the beginning of this mini stream, there is a mini \hi\ knot at $(1,~b)\approx(154\arcdeg,~33\arcdeg)$ that appears to be the start of a new Rayleigh-Taylor spike. At $(151\arcdeg,~35\arcdeg)$, there is a small H\textsc{~i} knot that branches off in the direction of Complex~A's motion, which would occur if this globular is flowing into a low pressure pocket behind core~AIII. Along this stream, there are two \hi\ knots at $(156\arcdeg,~35\arcdeg)$ and $(159\arcdeg,~36\arcdeg)$ that indicate that there are smaller Rayleigh-Taylor fingers forming off other fingers.

The gas associated with core~B (Figure~\ref{figure:decomp_region_B}) does not align with the main body of Complex~A (i.e., cores~A0--AVI; see Figure~\ref{figure:all_regions}). The gas that connects core~B to core~AVI has a relatively lower column density ($\log{\left(N_{\rm H\textsc{~i}}/\cm^2\right)}\lesssim18.5$) and is more diffuse compared to the A0--AVI cores. The entire core~B region likely represents a very large globule that is being swept away via ram-pressure stripping. However, this core region might be more diffuse than the other Rayleigh-Taylor fingers if it is being further disrupted by a turbulent wake that trails behind Complex~A as it travels through the Galactic halo.

\subsection{Kelvin-Helmholtz Instability Structures}\label{subsection:KH}

In addition to Rayleigh-Taylor Instabilities, Kelvin-Helmholtz Instabilities are also influencing Complex~A. As the outer layers of the complex ``rubs'' against the surrounding coronal gas, small tangential perturbations from shear-flow disturbances can form on its surface. If they become amplified, then some of affected gas will raise tangentially off the complex in the $\pm b_{\rm CA}$ directions. Elevated material can then be more easily swept away by the surrounding coronal gas through ram-pressure stripping as high pressure zones form on the leading edge of these structures and low pressure zones on the trailing edge. This elevated gas additionally can be influenced by Rayleigh-Taylor Instabilities in the direction of the host galaxy's gravitational potential well if it is able to maintain a gas density that is greater than the halo density and if it is not overpowered by ram-pressure stripping.

Kelvin-Helmholtz instabilities can affect all portions of the complex that are directly sliding against the surrounding coronal medium, but their signatures are more difficult to identify on the lower Galactic latitude half (or higher $b_{\rm CA}$ half) of Complex~A. This is because they are occurring in tandem with Rayleigh-Taylor instabilities and ram-pressure stripping, which have a stronger morphological impact on this HVC as evident by the numerous globules that hang from it (see Figure~\ref{figure:alonalat_coldens}). We therefore only identify Kelvin-Helmholtz instability structures on the higher latitude side of Complex~A. However, we stress that these Kelvin-Helmholtz instabilities could be exacerbating the Rayleigh-Taylor structures that form on the lower latitude edge of this complex. 

Three small Kelvin-Helmholtz structures branch off of cores~A0 and~A1, which are marked in Figures~\ref{figure:alonalat_coldens}. All of these structures point roughly perpendicularly off the surface of Complex~A with a slight tilt in the direction of Complex~A's motion. This is interesting as ram-pressure affects should cause these structures to tilt in the anti-direction of motion, but interactions with the denser gas near the Milky Way's disk may have affected their orientation. This unusual orientation is also shared by the low-latitude globular that hangs off of core~A0. As core~A0 is the leading core, its leading edge is being heated eroded away by direct interactions with denser material near the Milky Way's disk. These interactions assisted in the formation of the Rayleigh-Taylor globular at $(l,\,b)\approx(132\arcdeg,\,23\arcdeg)$ and the Kelvin-Helmholtz structure at $(133\arcdeg,\,27\arcdeg)$ (see Figure~\ref{figure:decomp_region_A0}). All three of the Kelvin-Helmholtz structures are attached to cores~A0 and~A1 are connected by a thin filament, indicating that they will soon detach and evaporate into the surrounding coronal medium (see Figures~\ref{figure:decomp_region_A0} and~\ref{figure:decomp_region_A1}). Additionally, the higher \hi\ column density sub-cores that have formed at the tips of these structures might indicate that these they are developing or will develop Rayeigh-Taylor fingers. 

Two high-latitude ``wings'' protrude from Complex~A (see Figure~\ref{figure:all_regions}), one between cores~AIII and~AIV at $(l,~b)\approx(147\arcdeg,~41\arcdeg)$ (see Figure~\ref{figure:decomp_region_W2}) and another off core~AVI at $(160\arcdeg,~44\arcdeg)$ (see Figure~\ref{figure:decomp_region_W1}). Because the stems of these wings extend perpendicularly off of Complex~A, this indicates that they were formed by Kelvin-Helmholtz instabilities. These structures subsequently became elongated due to interactions with the surrounding coronal as this HVC fell through the Galactic halo. Interestingly, sub-\hi\ cores have formed in the tips of these wings which may indicate that they are starting to form Rayleigh-Taylor fingers. The odd forward leaning morphology of these wings could be a result of buoyancy instabilities that are causing this higher density gas to fall faster toward the disk. However, in the case of wing~1, unseen eddies or a low pressure zone in the turbulent wake that lies behind wing~2 could also be causing this wing to curl forward (see Figure~\ref{figure:alonalat_coldens}).
 
In a hydrodynamical simulation of gas streams, \citet{2004ApJ...615..586M} found that wings can form as a result of evolving thermal and Kelvin-Helmholtz instabilities. They found that as the wings grow, they can curve in the direction of motion of the main cloud due to a combination of Rayleigh-Taylor instabilities and entanglement with vortices that formed in the surrounding coronal gas, which erode away the middle of the wing on its leading side.  The numerous cloud fragments that lie behind these wings indicates that there is substantial turbulent mixing behind them, presumably caused by a wake that follows these wings.

\section{Discussion}\label{section:discussion}

The HVCs that are infalling onto the Milky Way will generally need to travel for a tens to hundreds of million years to reach the Galactic disk as they typically lie $|z|\lesssim10~\kpc$ above or below the disk \citep{1999Natur.400..138V, 2001ApJS..136..463W, 2007ApJ...670L.113W, 2008ApJ...672..298W, 2006ApJ...638L..97T, 2008ApJ...684..364T, 2011MNRAS.415.1105S, 2015A&A...584L...6R} and are moving with speeds of $50\lesssim\rm |v_{z}|\lesssim200~\kms$ relative to the disk. While they are traversing the Galactic halo, they are gradually eroding away into the surrounding coronal medium. \citet{2009ApJ...698.1485H} and \citet{2007ApJ...670L.109B} predict that HVCs with $M_{\rm H\textsc{~i}} < 10^{4.5}~M_\odot$ will become fully ionized through Kelvin-Helmholtz instabilities within $\tau_{\rm KH}\lesssim100~{\rm Myr}$ and therefore they will not typically reach the Milky Way's disk. \citet{2011ApJ...739...30K} project that up to 70\% of the hydrogen in HVCs with masses of $M_{\rm H\textsc{~i}}\gtrsim10^5~M_\odot$ can remain neutral for a few hundred million years, which means that the large complexes are likely to survive their journey. However, higher mass HVCs that have a stream morphology, or that have a fractured surface are similarly vulnerable to rapid evaporation due to their increased surface area. Additionally, HVCs can become even more vulnerable to their surroundings if they become fragmented as a result of thermal instabilities (see \citealt{2004ApJ...615..586M}) or ``shock cascade'' processes (see \citealt{2007ApJ...670L.109B, 2015ApJ...813...94T}). 

While hydrodynamical instabilities assist in the destruction of HVCs, heat conduction \citep{2007A&A...472..141V, 2017MNRAS.470..114A}, self-gravity, and Magnetic fields \citep{1961hhs..book.....C, 2018ApJ...865...64G} are all processes that can suppress them (see \citealt{2012A&A...547A..43P} and \citealt{2018ApJ...865...64G}). As HVCs move through hot halo gas, they are being heated via thermal conduction, advection, and ionizing radiation, which means that these instabilities are at least partially suppressed due conduction. Although self-gravity would help stabilize HVCs, it is unlikely that these complexes are embedded within dark matter halos as their corresponding H\textsc{~i} Virial distances would place them millions of parsecs away \citep{1966BAN....18..421O, 2002ARA&A..40..487F}. 

The net effect that magnetic fields have on shaping HVCs is uncertain as hydrodynamic instabilities have been found to be mildly \citep{2016MNRAS.455.1309B} and strongly \citep{2015MNRAS.449....2M, 2016MNRAS.461..578G} suppressed and even enhanced \citep{2017ApJ...845...69G} in magnetohydrodynamic simulations (see \citealt{2018ApJ...865...64G}). It may be the case, however, that magnetic fields affect each kind of hydrodynamical instability differently. For instance, \citet{2016MNRAS.455.1309B} and \cite{2018ApJ...865...64G} found that magnetic fields inhibit Kelvin-Helmholtz instabilities, which helps to protect clouds against ablation by reducing their contact with halo material. In the case of thermal instabilities, \citet{2018MNRAS.476..852J} found that magnetic fields appear to promote thermal instabilities, which aids in their fragmentation. Similarly, \citet{1999ApJ...527L.113G}, \citet{2017ApJ...845...69G}, and \cite{2018ApJ...865...64G} found that magnetic fields enhanced Rayleigh-Taylor instabilities in the $z$~direction. Regardless, hydrodynamical effects dominate over magnetohydrodynamics in shaping clouds during most of their journey through the Galactic halo with the exception of when they are near the Galactic plane because the compressed leading edge of these clouds amplifies their magnetic field strength (see \citealt{2017ApJ...845...69G}). 

While there is uncertainty as to whether or not all HVCs have a magnetic field, the Smith Cloud  \citep{2013ApJ...777...55H}, Magellanic Bridge \citep{2017MNRAS.467.1776K}, and Leading Arm \citep{2010ApJ...725..275M} all have a detected magnetic field. However, as these three HVCs represent material that has been displaced from a galaxy (Smith Cloud from the MW: \citealt{2016ApJ...816L..11F}; Magellanic Bridge from the Magellanic Clouds: see \citealt{2012ARA&A..50..491P} for a review; Leading Arm from the SMC: \citealt{2018ApJ...854..142F}), these fields may have been inherited from their parent galaxy (Milky Way: \citealt{2015ASSL..407..483H}; Small Magellanic Cloud: \citealt{2015ApJ...806...94L}; Large Magellanic Cloud: \citealt{2012ApJ...759...25M}). It is unknown if the HVCs that originate from an intergalactic medium filament or though halo gas condensations will have a magnetic field.

No magnetic field has been directly measured for Complex~A. Nonetheless, \citet{2013ApJ...766..113V} placed indirect constraints on the strength of the  Complex~A's toroidal magnetic field by assuming that its broad \hi\ lines are a causes by a combination of thermal broadening and magnetic turbulence that results from Alfv\'en waves. In that study, they surmised that the lack of \ha\ emission detected in the Wisconsin \ha\ Mapper (WHAM) Northern Sky Survey (NSS) of the Milky Way \citep{2003ApJS..149..405H} was an indication that this complex is colder than $T_{\rm H\textsc{~i}}<7\times10^3~{\rm K}$ and therefore the \hi\ emission should only have narrow emission-line profiles of ${\rm FWHM}<25~\kms$. Assuming a distance of $d_{\odot}\approx200~{\pc}$ to Complex~A, they derived a field strength of $B\approx5~{\mu\rm G}$ with their model. However, there are two major issues with their assumptions: (1) The WHAM NSS does not span the kinematic extent of Complex~A (WHAM NSS: $-100\lesssim\vlsr\lesssim+100~\kms$), so no \ha\ emission from this complex would be present in this survey. \citet{2012ApJ...761..145B} mapped the \ha\ emission in Complex~A using the WHAM telescope and they detected \ha\ emission from the entire complex, which varied in strength from $30\lesssim I_{\rm H\alpha}\lesssim 100~{\rm mR}$. Therefore, broad \hi\ line profiles with $25\lesssim{\rm FWHM}\lesssim 35~\kms$ is not surprising as \ha\ emission peaks between 8,000$\lesssim T_{\rm H\alpha,\,peak}\lesssim$12,000~K. (2) Complex~A is much farther away than $d_\odot=200~\pc$.  \citet{2012ApJ...761..145B} also found that the inferred level of ionization based on the \ha\ emission could be produced by photoionization from the Milky Way and the extragalactic background if core~A0 lies roughly $6.3\lesssim d_{\odot} \lesssim 6.5~\kpc$, which is in agreement with distances derived via absorption-line studies (\citealt{1996ApJ...473..834W,1997MNRAS.289..986R, 1999Natur.400..138V, 2003ApJS..146....1W}). At this much larger distance, the \citet{2013ApJ...766..113V} model predicts that the magnetic-field strength would be on the same order as the external field at $B\lesssim1~{\mu\rm G}$.

While the predicting power of hydrodymanical simulations continue to improve, they are currently unable to fully resolve the detailed physics that are influencing HVCs. Smooth particle hydrodynamic simulations struggle to produce high resolution models that incorporate ram-pressure stripping, Kelvin-Helmholtz and Rayleigh-Taylor instabilities, turbulent motions, and magnetic fields (see \citealt{2015MNRAS.452.3853M}) as these codes can suppress entropy generation, underestimate vorticity generation, and impede efficient gas stripping \citep{2012MNRAS.424.2999S}. Adaptive mesh refinement codes have difficulty modeling diffusion as two mediums rub past each other at supersonic bulk velocities \citep{2015MNRAS.452.3853M}, which often leads to a suppression of shear instabilities as HVCs are moving through the Galactic halo at subsonic, transonic, and supersonic speeds  \citep{2011ApJ...739...30K}. Moving Voronoi mesh simulations can introduce noise on small spatial scales that are associated with mesh's motion \citep{2012MNRAS.423.2558B, 2013MNRAS.428.2840H}, which can result in second order instabilities in shear flows \citep{2015MNRAS.452.3853M}. Theoretical efforts are further hindered by having few observationally resolved examples of these processes to anchor their models.  

Until now, only one complete dataset of high resolution ($\Delta\theta\approx9\farcm1$) and high sensitivity ($\log{\left(\rm N_{H\textsc{~i}}/\cm^{-2}\right)\approx17.2}$) H\textsc{~i}~21-cm dataset exists for one of the Galaxy's HVC complexes: the Smith Cloud. This infalling complex is presently located at a Galactic height of $z\approx-3~\kpc$ and is estimated to impact with the Galactic disk in roughly $27~{\rm Myr}$ \citep{2008ApJ...679L..21L}.  The chemical composition of this HVC ($Z=0.53^{+0.21}_{-0.15}~Z_\odot$: \citealt{2016ApJ...816L..11F}) indicates that it likely originated from a Galactic fountain. Although the present mass of this complex of $\rm M_{\rm total
}\gtrsim2\times10^6~M_\odot$ (neutral: \citealt{2008ApJ...679L..21L}; ionized: \citealt{2009ApJ...703.1832H}) is much larger than anticipated from the energetic processes occurring within the Galactic disk \citep{2016ApJ...816L..11F}, hydrodynamical simulations suggest that when its high metallicity gas mixes with the surrounding coronal gas it can provide an avenue for the halo to cool and condense onto the complex (e.g., \citealt{2010MNRAS.404.1464M, 2013MNRAS.433.1634M, 2015MNRAS.447L..70F}. The cooled coronal gas can accrete onto the HVC, enabling it to grow as it travels through the halo \citep{2016MNRAS.462.4157A}. Because this  HVC has a measured average magnetic field strength along the line-of-sight (LOS) of $\vec{B}_{\rm LOS}\gtrsim8~\mu\rm G$ \citep{2013ApJ...777...55H}, it could be at least partially resistant to Kelvin-Helmholtz instabilities. Nonetheless, Rayleigh–Taylor instabilities have been observed in the \hi\ of this complex \citep{2019ApJ...871..215B}.

However, low metallicity halo clouds that are inefficient at cooling will instead more easily erode into the halo (e.g., \citealt{2012ApJ...745..148J}). HVC Complex~A is one such low metallicity cloud (${Z}=0.1~{Z}_\odot$: \citealt{1994A&A...282..709K, 1995A&A...302..364S,  1999Natur.400..138V, 2001ApJS..136..463W, 2012ApJ...761..145B}).  This HVC further has no detected magnetic field, so it is unknown if they are suppressing or enhancing hydrodynamic instabilities along its length. In this study, we have resolved the H\textsc{~i} morphology of this complex in unprecedented detail, which enable us to identify morphological structures that are associated with ram-pressure stripping and thermal, Rayleigh-Taylor, and Kelvin-Helmholtz instabilities. This study provides the first opportunity to trace all of these hydrodynamic instability signatures in a low metallicity HVC that may have originated from an intergalactic-medium filament.

Resolution matters. Previous observations of this gas stream using the Leiden/Argentine/Bonn (LAB) survey with a $\Delta\theta=0\fdg6$ resolution \citep{1997agnh.book.....H} severely spatially smoothed the emission so that only its bulk structure is resolved. \citet{1973A&AS...12..209G} and \citet{1976A&AS...23..181D} presented high angular resolution observations of Complex~A at $10\le\Delta\theta\le20\arcmin$, but they were much less sensitive at $0.5\lesssim{T_{\rm B,~1\sigma}}\lesssim1~\rm K$ and therefore these observations provided limited information as to how this complex is interacting with its environment. Although the \hi\ 4-PI (HI4PI) survey does have $\Delta\theta=16\farcm2$ resolution with a 1-sigma ${T_{\rm B,~1\sigma}}=43~{\rm mK}$ per $\Delta \rm v_{\rm bin}=1.29~\kms$ or 3-sigma $\log({N_{\rm H\textsc{~i},~3\sigma}}/\cm^{-2})=18.1$ sensitivity for a line with $\rm{FWHM}=20~\kms$ \citep{2016A&A...594A.116H}, the beam size of the GBT observation we are presenting in this study span an angular diameter that is a factor of $3.2\times$ smaller. At the angular resolution and sensitivity of the HI4PI survey, Rayleigh-Taylor instability fingers are difficult to identify (see Figure~3 of \citealt{2018MNRAS.474..289W}). Without high resolution observations of HVCs, like the ones presented this study, we will not be able to identify morphological and kinematical signatures of hydrodynamic instabilities, which are needed understand the survivability these complexes as they traverse the Galactic halo and anchor simulations.  

\section{Summary}\label{section:summary}
 
In this study, we explore the kinematics and morphology of the neutral hydrogen gas of Complex~A. We present a kinematically resolved H\textsc{~i}~21-cm map of this HVC over a $-230\le\rm v_{LSR}\le -90~\kms$ velocity range that spans a $600$-square degree area across the sky. This survey has a sensitivity of  $17.1\lesssim\log({N_{\rm H\textsc{~i},\,1\sigma}}/\cm^{-2})\lesssim17.9$ for lines with a ${\rm FWHM}=20~\kms$  width. We finish with the main conclusions of our study: 

\begin{enumerate}
\item{\bf Bulk Motion:} There is a Galactic standard of rest frame velocity gradient of $\Delta\rm v_{GSR}/\Delta L=25~\kms/{\kpc}$ along the $\Delta L\approx6.4~\kpc$ length of Complex~A. This corresponds to a deceleration rate of $\langle a\rangle_{\rm GSR}=55~{\rm km}/{\rm yr}^2$, which will place this complex at the Galactic plane in $\Delta t\lesssim70~{\rm Myrs}$.
\item{\bf Ram-Pressure Stripping:} Numerous H\textsc{~i} cloudlets along the Complex~A exhibit morphological signatures that are shaped by ram-pressure stripping. The cores~A0, AI, AII, and AIII tend to be compressed in the direction of motion with diffuse following gas (see  Figures~\ref{figure:all_regions}--\ref{figure:decomp_region_A3}). Much of the gas that extends off the low-latitude edge of the complex is tilted in the anti-direction of motion. This includes an ``H\textsc{~i} loop'' that extends off of core~A0, wispy gas that hangs from core~AII, a relatively multi-core filament and a small filament that branches off of core~AIV, and the entire core~B region (see Figure~\ref{figure:alonalat_coldens}).
\item{\bf Rayleigh-Taylor Instabilities:} We have identified numerous Rayleigh-Taylor fingers that hang from the lower latitude edge of the Complex~A stream. This includes a finger that hangs off core~A0 and curves upward in Complex~A's direction of motion (see Figure~\ref{figure:decomp_region_A0}), suggesting that this low-latitude gas is interacting with higher density gas near the Galactic disk. Fingers also extend off cores~AI, AII, AIII, and AIV (see Figures~\ref{figure:decomp_region_A1}, \ref{figure:decomp_region_A3}, and \ref{figure:decomp_region_A4}). The entire elongated and diffuse core~B region has the morphology of a large globular that branches off core~AVI and has been pushed backward due to ram-pressure stripping (see Figures~\ref{figure:all_regions} and~\ref{figure:decomp_region_B}). Additionally, the high density \hi\ sub-cores at the tip of the two high-latitude wings suggests that they could be forming globules.
\item{\bf Kelvin-Helmholtz instabilities:} Because both Rayleigh-Taylor instabilities and Kelvin-Helmholtz instabilities are simultaneously affecting the gas on the lower latitude edge of Complex~A, the Kelvin-Helmholtz signatures are difficult to isolate on this edge of the complex. On the high-latitude edge, there are two wings that extend tangentially from Complex~A that were formed through Kelvin-Helmholtz instabilities. After their initial formation, ram-pressure stripping elongated this gas and a combination of Rayleigh-Taylor instabilities and/or vortices in the surrounding coronal gas that are eroding caused them to curl slightly in the direction of motion. \\
\end{enumerate}

\acknowledgments

We thank the Research Apprentices Program at the Department of Physics and Astronomy at Texas Christian University, which enabled high school students to participate in this research. Barger received supported through NSF grant AST~1203059.  We thank Jay Lockman and Kevin Blagrave for providing us the deep Planck GBT \hi\ data cubes before they were publicly released.  We also thank Katie Chynoweth for sharing her deep M81 GBT \hi\ data cube with us.

\facility{Green Bank Telescope}
\software{GBTIDL and MPFIT}.

\bibliographystyle{aasjournal} 
\bibliography{References} 

\end{document}